\documentclass[entropy,article,accept,moreauthors,pdftex,12pt,a4paper]{mdpi_nologo}

\lastpage{2349}
\doinum{10.3390/e16042309}
\pubvolume{16}
\pubyear{2014}
\history{{\it Entropy} {\bf 16} (2014) 2309-2349. DOI:10.3390/e16042309.}

\usepackage{graphicx}
\usepackage{epsfig, graphics, epstopdf}
\usepackage{amssymb,amsmath}
\usepackage{subcaption}
\usepackage{soul,booktabs}
\allowdisplaybreaks

\newtheorem{definition}{Definition}
\newtheorem{lemma}{Lemma}
\newtheorem{theorem}{Theorem}

\def\ci{\perp\!\!\!\perp}

\Title{Measures of Causality in Complex Datasets with Application to Financial Data}

\Author{Anna Zaremba $^{1,}$* and Tomaso Aste $^{1,2}$ }

\address{%
{$^{1}$ Department of Computer Science, University College London, Gower Street, London WC1E 6BT, UK}\\
$^2$ Systemic Risk Centre, London School of Economics and Political Sciences, London WC2A2AE, UK }

\corres{E-Mail: A.Zaremba@cs.ucl.ac.uk; \linebreak Tel.: +44-0-20-3108-1305; Fax: +44-0-20-7387-1397.}

\abstract{This article investigates {the} causality structure of financial time series. We \linebreak concentrate on three main approaches to measuring causality: linear Granger causality, \linebreak kernel generalisations of Granger causality (based on ridge regression and {the} \linebreak Hilbert--Schmidt norm of the cross-covariance operator) and transfer entropy, examining each method and comparing their theoretical properties, with special attention given to the ability to capture nonlinear causality. We also present the theoretical benefits of applying non-symmetrical measures rather than symmetrical measures of dependence. We {apply} the measures to a range of simulated and real data. The simulated data sets {were} generated with linear and several types of nonlinear dependence, using bivariate, as well as multivariate {settings}. {An application} to real-world financial data highlights the practical difficulties, as well as the potential of the methods. We use two { real data sets}: (1) U.S. inflation and one-month Libor; (2) S$\&$P data and exchange rates for the following currencies: AUDJPY, CADJPY, NZDJPY, AUDCHF, CADCHF, NZDCHF. Overall, we {reach} the conclusion that no single method can be recognised as the best in all circumstances, and each of the methods has its domain of best applicability. We also {highlight} areas for improvement and\linebreak future research.
}


\begin{document}
\newpage


\section{Introduction}

Understanding the dependence between time series is crucial for virtually all complex systems studies, and the ability to describe the causality structure in financial data can be very beneficial to\linebreak financial institutions.

This paper concentrates on four measures of what could be referred to as ``statistical causality''. There is an important distinction between the ``intervention-based causality'' as introduced by\linebreak Pearl \cite{pearl_causality:_2000} and ``statistical causality'' as developed by Granger \cite{granger_testing_1980}. The first concept combines statistical and non-statistical data and allows one to answer questions, like ``if we give a drug to a patient, \textit{i.e.}, intervene, will {their chances of} survival increase?''. {Statistical} causality does not answer such questions, because it {does not} operate on the concept of intervention and only {involves} tools of data analysis. Therefore, the causality in a statistical sense is a type of dependence, where we infer direction as a result of the knowledge of temporal structure and the notion that the cause has to precede the effect. It can be useful for financial data, because it is commonly modelled as a single realisation of a stochastic process: a case where we cannot talk about intervention in the sense that is used by Pearl.

We {say} that $X$ causes $Y$ in the sense of statistical (Granger) causality if the future of $Y$ can be better explained with the past of $Y$ and $X$ rather than the past of $Y$ only. We will expand and further formalise this concept using different models.

To quote Pearl ``Behind every causal claim, there must lie some causal assumption that is not discernible from the joint distribution and, hence, not testable in observational studies'' (\cite{pearl_causality:_2000}, p 40). Pearl emphasises the need to clearly distinguish between the statistical and causal terminology, and while we {do not} follow his nomenclature, we agree that {it is} important to remember that statistical causality is not capable of discovering the ``true cause''.
Statistical causality can be thought of as a type of dependence, and some of the methods used for describing statistical causality derive from methods used for testing for independence.

The choice of the most useful method of describing causality has to be based on the characteristics of the data, and the more we know about the data, the better choice we can make. In the case of financial data, the biggest problems are the lack of stationarity and the presence of noise. If we also consider that the dependence is likely to exhibit nonlinearity, model selection becomes an important factor that needs to be better understood.

The goal of this paper is to provide a broad analysis of several of the existing methods to quantify causality. The paper is organised as follows:
 In {Section} \ref{methods}, we provide all the background information on the methods, as well as the literature review. In {Section} \ref{testing}, we describe 
 the practical aspects, the details of implementation, as well as the methodology of testing and the results of testing on synthetic data; financial and other 
 application are described {in Section} \ref{applications}. In {Section} \ref{discussion}, we provide a discussion of the methods, applications and perspectives. 
 Section \ref{conclusions} contains a brief summary. {Finally,} in \linebreak {Appendices} \ref{solving:rr}--\ref{proced:crossval}, we provide the {mathematical background and other 
 supplementary material.}

%
%
%
%
%

%
%

\newpage

\section{Methodology: Literature Review}
\label{methods}

\vspace{-12pt}
\subsection{Definitions of Causality, Methods}

The first mention of causality as a property that can be estimated {appeared} in 1956 in a paper by Wiener \cite{wiener_norbert_theory_1956}: ``For two simultaneously measured signals, if we can predict the first signal better by using the past information from the second one than by using the information without it, then we call the second signal causal to the first one.''

The first practical implementation of the concept was introduced by Clive Granger, the 2003 Nobel prize winner in {economics}, in {1963 \cite{granger_economic_1963} and 1969 \cite{granger_investigating_1969}}. The context in which Granger defined causality was that of linear autoregressive models of stochastic processes. Granger {described} the main properties that a cause should have: it should occur before the effect and it should contain unique information about the effect that is not contained in other variables. In his works, Granger included in-depth discussions of
 what causality means and how the statistical concept he introduced differed from deterministic causation.

\subsubsection{Granger Causality}

In the most general sense, we can say that a first signal causes \index{causality} a second signal if the second signal can be better predicted when the first signal is considered. It is Granger causality if the notion of time is introduced and the first signal precedes the second one. In the case when those two signals are simultaneous, we will use the term instantaneous coupling.

Expanding on the original idea of Granger, the two studies published by Geweke in 1982 \cite{geweke_measurement_1982} and in 1984 \citep{geweke_measures_1984} included the idea of feedback and instantaneous causality (instantaneous coupling). Geweke defined indices that measure the causality and instantaneous coupling with and without side information. While the indices introduced by Geweke are but one of a few alternatives that are used for quantifying Granger causality, these papers and the measures introduced therein are crucial for our treatment of causality. Geweke defined the measure of linear feedback, in place of the strength of causality used by Granger, which is one of the alternative Granger causality measures that is prevalent in the literature 
\cite{note1}.

We {use} notation and definitions that derive from \cite{amblard_kernelizing_2012}, { and we} generalise {them}. Let $\{ X_{t}\}, \{Y_{t}\}, \{Z_{t}\} $ be three stochastic processes. For any of the time series, using subscript $t$, as in $ X_{t}$, it will be understood as a random variable associated with the time $t$, while using superscript $t$, {as in} $ X^{t}$, it will be understood as the collection of random variables up to time $t$. {Accordingly, we use} $x_{t}$ and $x^{t}$ as realisations of those random variables.
\begin{definition}{(Granger causality)}
\label{def:Gcausality}\index{Granger causality}
$Y$ does not Granger cause $X$, relative to the side information, $Z$, if for all $ t \in \mathbb{Z} $:
\begin{equation}
P\left( X_{t} \mid X^{t-1}, Y^{t-k}, Z^{t-1} \right) =
P\left( X_{t} \mid X^{t-1},Z^{t-1} \right) ,
\end{equation}
where $k$ is any natural number and $P\left( \cdot \mid \cdot \right)$ stands for conditional probability distribution. If $k = 0$, we say that $Y$ does not instantaneously cause $X$ (instantaneous coupling):
\begin{equation}
P\left( X_{t} \mid X^{t-1}, Y^{t}, Z^{t-1} \right) =
P\left( X_{t} \mid X^{t-1},Z^{t-1} \right).
\end{equation}
\end{definition}


In the bivariate case, the side information, $ Z^{t-1} $, will simply be omitted. The proposed way of defining instantaneous coupling is practical to implement, but {it} is only one of several alternative definitions, none of which is \textit{a priori} {optimal}. Amblard \cite{amblard_relation_2012} recommends including $Z^{t}$ rather than $Z^{t-1}$ to ensure that the measure precludes confusing 
\cite{note2}
the instantaneous coupling of $X$ and $Y$ with that of $X$ and $Z$. Definition \ref{def:Gcausality} is very general and {does not} impose how the equality of the distributions should be assessed. The original Granger's formulation of causality is in terms of the variance of residuals for the least-squares predictor \citep{granger_investigating_1969}. There are many ways of testing that; here, we will, to a large degree, follow the approach from \cite{amblard_kernelizing_2012}.

Let us here start by introducing the measures of (Granger) causality that were originally proposed by Geweke in \citep{geweke_measurement_1982,geweke_measures_1984}. Let $\{ X_{t}\},\{ Y_{t}\}$ be two univariate stochastic processes and $\{ Z_{t}\}$ be a multivariate process (the setting can be generalised to include multivariate $ \{X_{t}\}, \{Y_{t}\}$).
We assume a vector autoregressive representation; hence, we assume that $\{X_{t}\}$ can be modelled in the following\linebreak  general way:
\begin{equation}
\label{eq:var:linf}
X_{t} = L_{X} (X^{t-1}) + L_{YX}(Y^{t-1}) + L_{ZX}(Z^{t-1}) + \varepsilon_{X,t}
\end{equation}
where {$L_{X}, L_{YX}, L_{ZX}$ are linear functions.} In Equation (\ref{eq:var:linf}), we allow some of the functions, $L_{\bullet}$, to be equal to zero everywhere. For example, if we fit $\{X_{t}\}$ with the model (\ref{eq:var:linf}) without any restrictions and with the same model, but adding the restrictions that $L_{YX}=0${:
\begin{equation}
\label{eq:var:linf2}
X_{t} = \tilde{L}_{X} (X^{t-1}) +\tilde{L}_{ZX}(Z^{t-1}) + \tilde{\varepsilon}_{X,t}
\end{equation}
then we can quantify the usefulness of including $\{Y_{t}\}$ in explaining $\{X_{t}\}$ with Geweke's measure of causality formulated as follows}:

\begin{equation}
\label{eq:var:Gew}
F_{Y\rightarrow X\parallel Z} = \log \dfrac{ var(X_{t} \mid X^{t-1}, Z^{t-1}) }{ var(X_{t} \mid X^{t-1}, Y^{t-1}, Z^{t-1}) } = \log \dfrac{ var(\tilde{\varepsilon}_{X,t}) }{ var(\varepsilon_{X,t}) }.
\end{equation}

Analogously, Geweke's measure of instantaneous causality (instantaneous coupling) {is} defined as:

\begin{equation}
\label{eq:var:Gewinst}
F_{Y\cdot X\parallel Z} = \log \dfrac{ var(X_{t} \mid X^{t-1}, Z^{t-1}) }{ var(X_{t} \mid X^{t-1}, Y^{t}, Z^{t-1}) } = \log \dfrac{ var(\tilde{\varepsilon}_{X,t}) }{ var(\hat{\varepsilon}_{X,t}) }, \
\end{equation}
where $\hat{\varepsilon}_{X,t}$ corresponds to yet another possible model, where both the past and present of $Y$ are considered: $X_{t} = \hat{L}_{X} (X^{t-1}) + \hat{L}_{YX}(Y^{t}) + \hat{L}_{ZX}(Z^{t-1}) + \hat{\varepsilon}_{X,t}$.

In a later section, we will present {a} generalisation of Geweke's measure using kernel methods.


\subsubsection{Kernels}
\label{kernels}

Building on Geweke's linear method of quantifying causality{, here, we} introduce a nonlinear measure that uses the ``kernel trick'', a method from machine learning{,} to generalize liner models.

There is a rich body of literature on causal inference from {the} machine learning perspective. Initially, the interest concentrated on testing for independence \cite{gretton_kernel_2005,gretton_measuring_2005,sun_kernel-based_2007,fukumizu_kernel_2008,gretton_kernel_2008}; but later, it was recognised that independence and non-causality are related, and the methods for testing one could be applied for testing the other \cite{sun_kernel-based_2007,guyon_causality:_2010}.

In {particular, in }the last several years, kernelisation has become a popular approach for generalising linear algorithms in many fields. The main idea {underlying} kernel methods is that nonlinear relationships between variables can become linear relationships between functions of the variables. This can be done by embedding (implicitly) the data into a Hilbert space and searching for a meaningful linear relationships in that space. The main requirement of kernel methods is that the data must not be represented individually, but only in terms of pairwise comparisons between the data points. {As} a function of two variables, {the} kernel function can be interpreted as a comparison function. It can also be {thought of} as a generalization of an inner product, such that the inner product is taken between functions of the variables; these functions are called ``feature maps''.

{In 2012, Amblard \textit{et al}. published the paper \citep{amblard_kernelizing_2012}, which}, to the best of our knowledge, is the first suggesting the generalisation of Granger causality using ridge regression. To some degree, this late development is surprising, as ridge regression is a well-established method for generalising linear regression and introducing kernels; it has a very clear interpretation, good computational properties and {a} straightforward way of optimising parameters.

An alternative approach to kernelising Granger causality {was} proposed by Xiaohai Sun in \cite{sun_assessing_2008}. Sun { proposed the} use of the square root of the Hilber--Schmidt norm of the so-called conditional \linebreak cross-covariance operator ({see Definition} \ref{def:crosscov}) in the feature space to measure the prediction error and {the use of a permutation test\index{permutation test} to quantify} the improvement of predictability. While {neither} of the two kernel approaches described in this paper is based on Sun's article, they are closely related. In particular, the concept of the Hilbert--Schmidt Normalised Conditional Independence Criterion (HSNCIC) \index{Hilbert--Schmidt Normalised Conditional Independence Criterion (HSNCIC)} {is from} a similar family as the one explored by Sun. {Another method of kernelising Granger causality has been described by Marinazzo \textit{et al}. \cite{marinazzo_kernel-granger_2008}.} Below, we are following the approach from \cite{schoelkopf_kernel_2004}. Please refer to the Appendix \ref{def:fa} for supplementary information {on} functional analysis and Hilbert spaces.

Let us denote with $S = (x_{1},..., x_{n})$ a set of n observations from the process, $\{X_{t}\}$. We suppose that each observation, $x_{i}$, is an element of some set, $ \mathcal{X}$. To analyse the data, and use the ``kernel trick'': we create a representation of the data set, $S$, that uses pairwise comparisons $k:\mathcal{X}\times \mathcal{X} \rightarrow \mathbb{R}$ of the points of the set, $S$, rather that the individual points. The set, $S$, is then represented by $n \times n$ comparisons $k_{i,j} = k(x_{i}, x_{j})$.

\begin{definition}{(Positive definite kernel)} A function $k: \mathcal{X} \times \mathcal{X} \rightarrow \mathbb{R}$ is called a positive definite kernel \index{kernel (positive definite)} {if and only if} it is symmetric, that is, $ \forall x,x' \in \mathcal{X}, k(x,x') = k(x',x)$ and {positive (semi-) definite}, that is:
\begin{equation}
\forall x_{1}, ..., x_{n} \in X \; \; \; \forall c_{1}, ..., c_{n} \in \mathbb{R} \; \; \; \sum_{i=1}^{n}\sum_{j=1}^{n}c_{i}c_{j}k(x_{i},x_{j})\geqslant 0 .
\end{equation}
\end{definition}
We will use the name kernel instead of the positive (semi-) definite kernel henceforth.

\begin{theorem}
For any kernel, $k$, on space $\mathcal{X}$, there {exists} a Hilbert space, $F$, and a mapping $\phi:\mathcal{X} \rightarrow F$, such that \cite{schoelkopf_kernel_2004}:
\begin{equation}
k(x,x') = \langle \phi(x), \phi(x') \rangle, \; \; \; \mbox{for any } x, x' \in \mathcal{X},
\end{equation}
where $\langle u, v \rangle, u, v \in \mathcal{F}$ represents an inner product in $ \mathcal{F}$.
\end{theorem}

The above theorem leads to an alternative way of defining a kernel. It shows how we can create a kernel provided we have a feature map. Because the simplest feature map is {an identity} map, this theorem proves that the inner product is a kernel.

The kernel trick \index{kernel trick}is a simple and general principle based on the fact that kernels can be thought of as inner products. {The kernel trick} can be stated as follows \cite{schoelkopf_kernel_2004}: ``Any algorithm for vectorial data that can be expressed only in terms of dot products between vectors can be performed implicitly in the feature space associated with any kernel, by replacing each dot product by a kernel evaluation.''

In the following two sections, we will illustrate the use of the kernel trick in two applications: (1) the extension to the nonlinear case of linear regression-Granger causality; and (2) the reformulation of concepts, such as covariance and partial correlations, to the nonlinear case.

\subsubsection{Kernelisation of Geweke's Measure of Causality}
\label{kernel:Gcausality}

Here, we will show how the theory of reproducing kernel Hilbert spaces can be applied to generalise the linear measures of Granger causality as proposed by Geweke. {To do so}, we will use the standard theory of ridge regression.

First of all, let us go back to the model formulation of the problem that we had before (Equation \ref{eq:var:linf}). We assumed that $ \{X_{t}\}, \{Y_{t}\} $ are two univariate stochastic processes and $ \{Z_{t}\} $ is a multivariate stochastic process. Let us {now} assume to have a set of observations $S = \lbrace (x_{1}, y_{1}, z_{1}) , ..., (x_{n}, y_{n},z_{n})\rbrace, \; x_{t}, y_{t} \in \mathbb{R}, z_{t} \in \mathbb{R}^{k}$. The goal is to {find the best linear fit function, $f$, which describes the
given dataset, $S$
. The idea is similar to the linear Geweke's measure: to infer causality from the comparison of alternative models. In particular, four alternative functional relations between points are modelled: (1) between $x_t$ and its own past; (2) between $x_t$ and the past of $x_t, y_t$; (3) between $x_t$ and the past of $x_t, z_t$; and (4) between $x_t$ and the past of $x_t, y_t, z_t$.
In order to have a uniform notation for all four models, a new variable, $w$, is introduced, with $w_{i}$ symbolising either $x_{i}$ itself, or $(x_{i},y_{i})$, or $(x_{i},z_{i})$ or $(x_{i}, y_{i},z_{i})$.
Thus, the functional relationship between the data can be written as follows: for all $t$,} $x_{t} \simeq f(w_{t-1}^{t-p})$,
where $w_{t-1}^{t-p}$ is a collection of samples $w_{i}$ made from $p$ lags prior to time $t$, such that $w_{t-1}^{t-p} = (w_{t-p}, w_{t-p+1}, \ldots, w_{t-1})$.
For instance, in the case where $w$ represents all three time series: $w_{t-1}^{t-p} = (x_{t-p}, y_{t-p}, z_{t-p}, x_{t-p+1}, y_{t-p+1}, z_{t-p+1}, \ldots, x_{t-1}, y_{t-1}, z_{t-1})$.
In general, {$p$ could represent an infinite lag}, but for any practical case, it is reasonable to assume a finite lag and, therefore, $w_{t-1}^{t-p}\in \mathbb{X}$, where typically $\mathbb{X} = \mathbb{R}^{d}$ for $d = p$ if $w = x$, or $d = 2p$ if $w = (x,y)$ {or} $d = 2p + k p$ if $w = (x,y,z)$. {Least} squares regression (as in the {linear Granger causality discussed} earlier) involves looking for a real valued weight vector, $\beta$, such that $x_{t} \simeq \hat{x_{t}} = (w_{t-1}^{t-p})^{T} \beta$, \textit{i.e.}, choosing the weight vector, $\beta$, that minimises the squared error. The dimensionality of $\beta$ depends on the dimensionality of $w${; it is} a scalar in the simplest case of $w = x$, with $x$ being univariate.

It is well known that the {drawbacks of least squares regression include: poor effects with small sample size, no solution when data are linearly independent and overfitting.} Those problems can be addressed by adding to the cost function an additional cost penalizing {excessively large} weights of the coefficients. This cost, called the regulariser \cite{hastie_elements_2009} or regularisation term, introduces a trade-off between the mean squared error and a {squared} norm of the weight vector. The regularised cost function is now:

\begin{equation}
\label{eq:reg:costf}
\beta^{\ast} = \operatornamewithlimits{argmin}_{\beta} \dfrac{1}{m} \sum_{i=p+1}^{n}( (w_{i-1}^{i-p})^{T} \beta - x_{i})^2 + \gamma \beta^{T}\beta,
\end{equation}
with $m=n-p$ for a {more concise} notation.

Analogously to the least squares regression weights, the solution of ridge regression (obtained in Appendix \ref{solving:rr}) can be written in the form of primal weights $\beta^{\ast}$:
\begin{equation}
\beta^{\ast} = (\mathbf{W}^{T}\mathbf{W}+ \gamma m I_{m})^{-1}\mathbf{W}^{T}x,
\end{equation}
{where we use} the matrix notation $\mathbf{W} = ((w_{p}^{1})^{T}, (w_{p +1}^{2})^{T}, ..., (w_{n-1}^{n-p})^{T})^{T}$, or in other words, a matrix with {rows} $w_{p}^{1}, w_{p +1}^{2}, ..., w_{n-1}^{n-p}$; $x = (x_{p+1}, x_{2,t}, ..., x_{n})^{T}$; {and} $I_{m}$ denotes an identity matrix of size $m\times m$.

However, we want to be able to apply kernel methods, which require that the data is represented in the form of inner products rather than the individual data points. As is explained in Appendix \ref{solving:rr}, the weights, $\beta$, can be represented as a linear combination of the data points: $\beta = \mathbf{W}^{T} \alpha$, for some $\alpha$. This second representation results in the dual solution, $\alpha^{\ast}$, that can be written in terms of $\mathbf{W}\mathbf{W}^{T}$ and that depends on the regulariser, $\gamma$:
\begin{equation}
\label{eq:dual:sol}
\alpha^{\ast} = (\mathbf{W}\mathbf{W}^{T} + \gamma m I_{m})^{-1} x,
\end{equation}

This is where we can apply the kernel trick that will allow us to introduce kernels to the regression setting above.
{For this purpose, we} introduce kernel similarity function $k$, which we apply to elements of $\mathbf{W}$. {The Gram matrix built from evaluations of kernel functions on each row of $\mathbf{W}$ is denoted by $\mathbf{K_{w}}$}:
\begin{equation}
(\mathbf{K_{w}})_{i,j} = k(w_{i-1}^{i-p},w_{j-1}^{j-p}), \; \; \; \mbox{for }i,j = p+1 \cdots n.
\end{equation}

The kernel function, $k$, has the associated linear operator $k_{w} = k(\cdot,w)$. {The} representer theorem (Appendix \ref{def:fa}) allows us to represent the result of our minimisation (\ref{eq:reg:costf}) as a linear combination of kernel operators \cite{amblard_kernelizing_2012}. The optimal prediction can now be written in terms of the dual weights in the\linebreak following way:

\begin{equation}
\label{eq:dual:pred}
\hat{x_{t}} = k_{w}(w_{t-1}^{t-p})^{T}(\mathbf{K_{w}} + \gamma m I_{m})^{-1} x.
\end{equation}

The mean square prediction error can be calculated by averaging over the whole set of realisations:

\begin{equation}
\label{eq:dualerror}
var_K(X_{t} \mid W^{t-1}) = \dfrac{1}{m} \sum_{j=1}^{l}(x_{j} - \hat{x_{j}})^{2} = \dfrac{1}{m} (\mathbf{K_{w}} \alpha^{\ast} - x)^{T}(\mathbf{K_{w}} \alpha^{\ast} - x),
\end{equation}
where $\hat{x_{j}}$ denotes a fitted value of $x_{j}$.

Analogously to the Geweke's indices from Equations (\ref{eq:var:Gew}), we {now} define {kernelised} Geweke's indices for causality and instantaneous coupling using the above framework:
\begin{equation}
\label{eq:ker:Gew}
\begin{split}
G_{Y\rightarrow X\parallel Z} &= \log \dfrac{ var_K (X_{t} \mid X^{t-1}, Z^{t-1}) }{ var_K (X_{t} \mid X^{t-1}, Y^{t-1}, Z^{t-1}) }\\
G_{Y\cdot X\parallel Z} &= \log \dfrac{ var_K (X_{t} \mid X^{t-1}, Z^{t-1}) }{ var_K(X_{t} \mid X^{t-1}, Y^{t}, Z^{t-1}) },
\end{split}
\end{equation}
extending in this way Geweke's measure of causality to the non-linear case.


\subsubsection{Hilbert--Schmidt Normalized Conditional Independence Criterion}\index{Hilbert--Schmidt Normalised Conditional Independence Criterion (HSNCIC)}
\label{HSNCIC}

Covariance can be used to analyse second order dependence{,} and in the special case of variables with Gaussian distributions{,} zero covariance is equivalent to independence. In 1959, Renyi \cite{renyi_measures_1959} pointed out that to assess independence between random variables $X$ and $Y$, one can use maximum correlation $S$ defined as follows:
\begin{equation}
S(X, Y) = \sup_{f,g} \left( corr(f( X), g( Y)) \right)
\end{equation}
where $f$ and $g$ are any Borel-measurable functions for which $f(X)$ and $g(Y)$ have finite and positive variance. Maximum correlation has all of the properties that Renyi postulated for an appropriate measure of dependence; most importantly, it equals zero if and only if the variables, $X$ and $Y$, are independent. However, the concept of maximum correlation is not practical, {as} there might not even exist such functions, $f_{0}$ and $g_{0}$, for which the maximum can be attained \cite{renyi_measures_1959}. Nevertheless, this concept has been used as a foundation of some kernel-based methods for dependence, such as kernel constrained covariance \cite{gretton_kernel_2005-1}.

This section requires some background from functional analysis and machine learning. For completeness, the definitions of the Hilbert--Schmidt norm and operator, tensor product and mean element are given in the Appendix \ref{def:fa} and follow \cite{gretton_measuring_2005,fukumizu_kernel_2008}.

{The cross-covariance operator is analogous to a} covariance matrix, but is defined for feature maps.
\begin{definition}{(Cross-covariance operator)}
\label{def:crosscov}\index{cross-covariance operator}
The cross-covariance operator is a linear operator $\Sigma_{XY}:\mathcal{H_{Y}} \rightarrow \mathcal{H_{X}}$ associated with the joint measure, $P_{XY}$, defined as:
\begin{equation}
\label{eq:Cxy:def}
\Sigma_{XY}:= \textbf{E}_{XY}[(\phi(X) - \mu_{X})\otimes (\phi(Y) - \mu_{Y})] =
\textbf{E}_{XY}[\phi(X) \otimes \phi(Y) ] - \mu_{X} \otimes \mu_{Y}
\end{equation}
\end{definition}
where we use symbol $\otimes$ for tensor product and $\mu$ for mean embedding (definitions in Appendix \ref{def:fa}). {The cross}-covariance operator applied to two elements of $\mathcal{H_{X}}$ and $\mathcal{H_{Y}}$ gives the covariance:
\begin{equation}
\begin{split}
\label{eq:Cxy:cov}
\langle f,\Sigma_{XY} g\rangle_{\mathcal{H_{X}}} = Cov(f(X),g(Y))
\end{split}
\end{equation}

The notation and assumptions follow
\cite{gretton_measuring_2005,sun_assessing_2008}: $\mathcal{H_{X}}$ denotes the reproducing kernel Hilbert space (RKHS) induced by a strictly positive kernel $k_{\mathcal{X}} : \mathcal{X} \times \mathcal{X} \rightarrow \mathbb{R}$, analogously for $\mathcal{H_{Y}}$ and $k_{\mathcal{Y}}$. $X$ is a random variable on $\mathcal{X}$; $Y$ is a random variable on $\mathcal{Y}$, and $(X,Y)$ is a random vector on $\mathcal{X} \times \mathcal{Y}$. We assume $\mathcal{X}$ and $\mathcal{Y}$ to be topological spaces, and measurability is defined with respect to the {relevant} $\sigma-$fields. The marginal distributions are denoted by $P_{X}, P_{Y}$ and the joint distribution of $(X,Y)$ by $P_{XY}$. The expectations, $\textbf{E}_{X}$, $\textbf{E}_{Y}$ and $\textbf{E}_{XY}$, denote the expectations over $P_{X}$, $P_{Y}$ and $P_{XY}$, respectively. To ensure $\mathcal{H_{X}}, \mathcal{H_{Y}}$ are included in, respectively, $L^{2}(P_{X})$ and $L^{2}(P_{Y})$, we consider only random vectors $(X,Y)$, {such that} the expectations, $\textbf{E}_{X}[k_{\mathcal{X}}(X,X)]$ and $\textbf{E}_{Y}[k_{\mathcal{Y}}(Y,Y)]$, are finite.

Just as {the} cross-covariance operator is related to the covariance, we can define an operator that is related to partial correlation:

\begin{definition}{(Normalised conditional cross-covariance operator \protect{\cite{fukumizu_kernel_2008}})}\index{normalised conditional cross-covariance operator} Using the cross-covariance operators, we can define the normalised conditional cross-covariance operator in the following way:
\begin{equation}
\label{eq:Vyxz}
V_{XY\mid Z} = \Sigma_{XX}^{-1/2} ( \Sigma_{XY} - \Sigma_{XZ}\Sigma_{ZZ}^{-1/2} \Sigma_{ZY} ) \Sigma_{YY}^{-1/2}
\end{equation}
\end{definition}

Gretton \textit{et al}. \citep{gretton_measuring_2005} state that for rich enough RKHS (\textcolor{black}{by ``rich enough'', we mean universal, \textit{i.e.}, dense in the sense of continuous functions on $\mathcal{X}$ with the supremum norm \cite{hofmann_kernel_2008})}, the zero norm of the cross-covariance operator {is equivalent to} independence, which can be written as:

\begin{equation}
\label{eq:crosscov:indep}
X \ci Y \Longleftrightarrow \Sigma_{XY} = 0
\end{equation}
where zero denotes a null operator. This equivalence is the premise from which follows {the use} of the Hilbert--Schmidt independence criterion (HSIC) as a measure of independence {(refer} to Appendix \ref{HSIC} for the information about HSIC).

It {was} shown in \cite{fukumizu_kernel_2008} that there is a relationship similar to (\ref{eq:crosscov:indep}) between the normalised conditional cross-covariance operator and conditional independence, which can be written as:
\begin{equation}
X \ci Y \mid Z \Longleftrightarrow V_{(XZ)(YZ)\mid Z} = 0
\end{equation}
where by $(YZ)$ and $(XZ)$, we denote extended variables. Therefore, the Hilbert--Schmidt norm of the conditional cross-covariance operator has been suggested as a measure of conditional independence. Using the normalised version of the operator has the advantage that it is less influenced by the marginals than {the} non-normalised operator, while retaining all the information about dependence. This is {analogous} to the difference between correlation and covariance.

\begin{definition}{(Hilbert--Schmidt {normalised conditional independence criterion} (HSNCIC)} We define the HSNCIC as the squared Hilbert--Schmidt norm of the normalised conditional cross-covariance operator, $V_{(XZ)(YZ)\mid Z}$:
\begin{equation}
\label{eq:HSNCIC}
HSNCIC:= \Vert V_{(XZ)(YZ)\mid Z}\Vert_{HS}^{2}
\end{equation}
where $\Vert \cdot \Vert_{HS}$ denotes Hilbert--Schmidt norm of an operator, defined in the {Appendix} \ref{def:fa}.
\end{definition}

For the sample $S = \lbrace (x_{1}, y_{1}, z_{1}) , ..., (x_{n}, y_{n},z_{n})\rbrace$, HSNCIC has an estimator that is both straightforward and has good convergence behaviour \citep{fukumizu_kernel_2008,seth_assessing_2012}. As shown {in Appendix} \ref{HSNCIC:est}, it can be obtained by defining empirical estimates of all of the components in following steps: first define mean elements $\hat{m}_{X}^{(n)}$ and $\hat{m}_{Y}^{(n)}$ and use them to define empirical cross-covariance operator $\hat{\Sigma}_{XY}^{(n)}$. Subsequently, using $\hat{\Sigma}_{XY}^{(n)}$, together with $\hat{\Sigma}_{XX}^{(n)}$ and $\hat{\Sigma}_{YY}^{(n)}$ obtained in the same way,
define $\hat{V}_{XY}^{(n)}$ for the empirical normalised cross-covariance operator. Note that $V_{XY}$ requires inverting $\Sigma_{YY}$ and $\Sigma_{XX}$; hence, to ensure invertibility a regulariser, $n \lambda I_{n}$, is added. The next step is to construct the estimator, $\hat{V}_{XY | Z}^{(n)}$, from $\hat{V}_{XY}^{(n)}$, $\hat{V}_{XZ}^{(n)}$ and $\hat{V}_{ZY}^{(n)}$. Finally, construct the estimator of the Hilbert--Schmidt norm of $\hat{V}_{ZY}^{(n)}$ as follows:
\begin{equation}
HSNCIC_{n}:= Tr[R_{(XZ)}R_{(YZ)} - 2R_{(XZ)}R_{(YZ)}R_{Z} + R_{(XZ)}R_{Z}R_{(YZ)}R_{Z}]
\end{equation}
where $Tr$ denotes a trace of a matrix and $R_{U} = K_{U}(K_{U} + n\lambda I)^{-1}$ and $K_{U}(i,j) = k(u_{i}, u_{j})$ is a Gram matrix. This estimator depends on the regularisation parameter, $\lambda$, which, in turn, depends on the sample size. Regularisation becomes necessary when inverting finite rank operators.



\subsubsection{Transfer Entropy} \index{transfer entropy}

Let us now introduce an alternative nonlinear information-theoretic measure of causality, which is widely used and provides us with an independent comparison for the previous methods.

In 2000, Schreiber suggested measuring causality as an information transfer, in the sense of information theory. He called this measure ``transfer entropy'' \cite{schreiber_measuring_2000}\index{transfer entropy}.
{Transfer entropy} has become popular among physicists and biologists, and there is a large body of literature on {the} application of transfer entropy to neuroscience. We refer to \cite{lindner_trentool:_2011} for a description of one of the best developed toolboxes for estimating transfer entropy. A comparison of transfer entropy and other methods to measure causality in bivariate time series, including extended Granger causality, nonlinear Granger causality, predictability improvement and two similarity indices, {was} presented
by Max Lungarella \textit{et al}. in \citep{lungarella_methods_2007}. A particularly exhaustive review of the relation between Granger causality and directed information {is presented by} Amblard \textit{et al}. \citep{amblard_relation_2012}, while for a treatment of the topic from the network theory perspective, refer to Amblard and Michel \citep{amblard_directed_2011}.

Transfer entropy {was} designed to measure the departure from the generalized Markov property stating that $P\left( X_{t} \mid X^{t-1}, Y^{t-1} \right) = P\left( X_{t} \mid X^{t-1} \right)$. From the definition of Granger causality (\ref{def:Gcausality}) for the bivariate case, \textit{i.e.}, with omitted side information $\{Z_{t}\}$, we can see that Granger non-causality should imply zero transfer entropy (proved by Barnett \textit{et al}. \cite{barnett_granger_2009} for the linear dependence of Gaussian variables and for Geweke's formulation of Granger causality).

Transfer entropy is related to and can be decomposed in terms of Shannon entropy, as well as in terms of Shannon mutual information:
\begin{definition}{(Mutual information)} Assume that $U,V$
 are discrete random variables with probability distributions $P_U(u_{i}), P_V(v_{i})$ and joint distribution $P_{UV}(u_{i},v_{i})$. Then, the {mutual information}, $I(U,V)$, is defined as:
\begin{equation}
I(U,V) = \sum_{i,j} P_U(u_{i},v_{j}) \log \dfrac{P_{UV}(u_{i},v_{j})}{P_U (u_{i})P_V(v_{j})} = H(U) - H(U \mid V)
\end{equation}
\end{definition}
with $H(U)=- \sum_{i} P_U(u_{i}) \log P_U(u_{i})$ the Shannon entropy and $H(U|V)= \sum_{i,j} P_{UV}(u_{i},v_{j}) \log \dfrac{P_V(v_{j})}{P_{UV} (u_{i},v_{j})}$ the Shannon conditional entropy.

For independent random variables, the mutual information is zero. Therefore, the interpretation of mutual information is that it can quantify the lack of independence between random variables, and what is particularly appealing is that it does so in a nonlinear way. However, being a symmetrical measure, mutual information cannot provide any information about the direction of dependence. A natural extension of mutual information to include directional information is transfer entropy. According to Schreiber, the family of Shannon entropy measures are properties of static probability distributions, while transfer entropy is a generalisation to more than one system and is defined in terms of transition\linebreak probabilities \citep{schreiber_measuring_2000}.

We assume that $X$, $Y$ are random variables. As previously, $X_{t}$ stands for a value at point $t$ and $X^{t}$ for a collection of values up to point $t$.

\begin{definition}{(Transfer entropy)}\index{transfer entropy} The transfer entropy $T_{Y \rightarrow X}$ is defined as:
\begin{equation}
T_{Y \rightarrow X} = H(X_{t}\mid X^{t-1}) - H(X_{t} \mid X^{t-1}, Y^{t-1})
\end{equation}
\end{definition}

Transfer entropy can be {generalised} for a multivariate system, for example \cite{barnett_granger_2009} defines conditional transfer entropy $T_{Y \rightarrow X \mid Z} = H(X_t\mid X^{t}, Z^{t}) - H(X_t \mid X^{t}, Y^{t}, Z^{t})$. {In this paper, we will calculate} transfer entropy only in the case of two variables. This is because the calculations already involve the estimation of the joint distribution of three variables ($X_t, X^{t}, Y^{t}$), and estimating the joint distribution of more variables would be impractical for time series of {the} length that we work with in\linebreak financial applications.

\section{Testing}
\label{testing}
\vspace{-12pt}

\subsection{Permutation Tests}
\label{permtest}

{Let us, first of all, emphasise} that in the general case, the causality measures introduced before should not be used as absolute values, but rather serve the purpose of comparison. While we observe that, on average, increasing the strength of coupling increases the value of causality, there is a large deviation in the results unless the data has been generated with linear dependence and small noise. Consequently, we need a way of assessing the significance of the measure as a way of assessing the significance of the causal relationship itself.
To achieve this goal, we shall use permutation tests, following the \linebreak approach in \cite{amblard_kernelizing_2012,sun_assessing_2008,seth_assessing_2012}.

By permutation test, we mean a type of statistical significance test in which we use random permutations to obtain the distribution of the test statistic under the null hypothesis. We would like to compare the value of our causality measure on the analysed data and on ``random'' data and conclude that the former is significantly higher.
We expect that destroying the time ordering should also destroy any potential causal effect, since statistical causality relies on the notion of time. Therefore, we create the distribution of $H_{0}$ by reshuffling $y$, while keeping the order of $x$ and $z$ intact. More precisely, let ${\pi_{1}, ..., \pi_{n_{r}}}$ be a set of random permutations. Then, instead of $y_{t}$, we consider $y_{\pi_{j}(t)}$, obtaining a set of measurements $G_{Y_{\pi_{j}} \rightarrow X \mid \mid Z}$ that can be used as an estimator of the null hypothesis $G_{Y \rightarrow X \mid \mid Z}^{0}$. We will accept the hypothesis of causality only if, for most of the permutations, the value of the causality measure obtained on the shuffled (surrogate) data is smaller than the value of causality measure of original data. This is quantified with {a \emph{p}-value} defined as follows:


\begin{equation}
\label{eq:accept}
p = \dfrac{1}{n_{r}}\sum_{j=1}^{n_{r}} \mathbf{1}( G_{Y_{\pi_{j}} \rightarrow X \mid \mid Z} > G_{Y \rightarrow X \mid \mid Z} )
\end{equation}
Depending on the number of permutations used, we suggest to accept the hypothesis of causality for the level of significance equal to $0.05$ or $0.01$. In our experiments, we {report} either single \emph{p}-values or sets of \emph{p}-values for overlapping moving windows. The latter is particularly useful when analysing noisy and non-stationary data. In the cases where not much data is available, we {do not} believe that using any kind of subsampling (as proposed by \cite{sun_assessing_2008,amblard_kernelizing_2012,seth_assessing_2012}) will be beneficial, as far as the power of the tests is concerned.

\subsection{Testing on Simulated Data}
\label{testing:simdata}
\vspace{-12pt}
\subsubsection{Linear Bivariate Example}
\label{lin:experiment}

\begin{figure}[H]
 \centering
 \includegraphics[width=5cm]{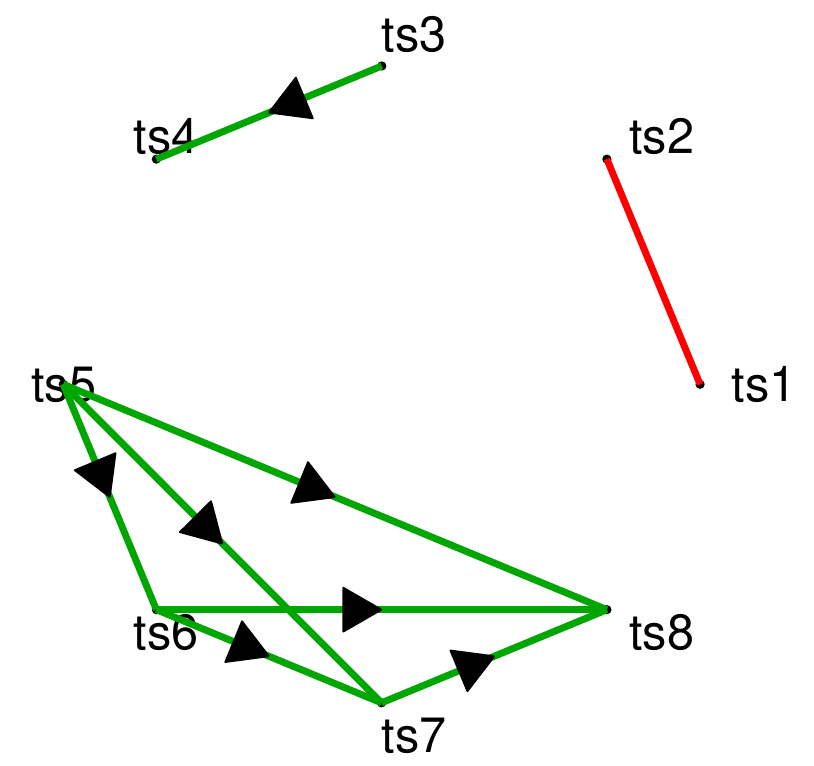}
 \caption{The directionality of causality between the eight simulated time series. Green lines represent causality, with the arrowheads indicating direction; the red line indicates instantaneous~coupling.}
 \label{fig:causstruct}
\end{figure}

Before applying the methods to real-world data, it is prudent to verify {whether they work} for data with known and simple dependence structure. We tested the methods on a data set containing eight time series with a relatively simple causal structure at different lags and some instantaneous coupling. We used the four methods to try to capture the dependence structure, as well as to figure out which lags show dependence. The data {was} simulated by first generating a set of eight time series from a Gaussian distribution with correlation matrix represented in {Table} \ref{tab:corr}. Subsequently, some of the series {were} shifted by one, two or three time steps to obtain the following ``causal'' relations: $x_{1} \longleftrightarrow x_{2}$ at Lag $0$, \textit{i.e.}, instantaneous coupling of the two variables, $x_{3} \rightarrow x_{4}$ at Lag 1, $x_{5} \rightarrow x_{6}$ at Lag 1, $x_{5} \rightarrow x_{7}$ at Lag 2, $x_{5} \rightarrow x_{8}$ at Lag 3, $x_{6} \rightarrow x_{7}$ at Lag 1, $x_{6} \rightarrow x_{8}$ at Lag 2, $x_{7} \rightarrow x_{8}$ at Lag 1. The network structure is shown in {Figure} \ref{fig:causstruct}, while the lags at which the causality occurs are given in the {Table} \ref{tab:lags}. {The length of the data is 250. }

\begin{table}[H]
\caption{The dependence structure of the simulated data. (\textbf{a}) The correlation matrix that has been used to generate the testing data; (\textbf{b}) Lags at which true dependence occurs, with the interpretation that the column variable causes the row variable.}
	\begin{subtable}{.5\linewidth}

		\begin{tabular}{c c c c c c c c c}
\toprule
		 & \textbf{ts
		1} & \textbf{ts2} & \textbf{ts3} & \textbf{ts4} & \textbf{ts5} & \textbf{ts6} & \textbf{ts7} & \textbf{ts8} \\
		\midrule
		ts1 & 1 & 0.7 & 0.1 & 0.1 & 0.1 & 0.1 & 0.1 & 0.1 \\
		ts2 & 0.7 & 1 & 0.1 & 0.1 & 0.1 & 0.1 & 0.1 & 0.1 \\
		ts3 & 0.1 & 0.1 & 1 & 0.7 & 0.1 & 0.1 & 0.1 & 0.1 \\
		ts4 & 0.1 & 0.1 & 0.7 & 1 & 0.1 & 0.1 & 0.1 & 0.1 \\
		ts5 & 0.1 & 0.1 & 0.1 & 0.1 & 1 & 0.7 & 0.7 & 0.7 \\
		ts6 & 0.1 & 0.1 & 0.1 & 0.1 & 0.7 & 1 & 0.7 & 0.7 \\
		ts7 & 0.1 & 0.1 & 0.1 & 0.1 & 0.7 & 0.7 & 1 & 0.7 \\
		ts8 & 0.1 & 0.1 & 0.1 & 0.1 & 0.7 & 0.7 & 0.7 & 1 \\
		\bottomrule
		\end{tabular}
\caption{}
\label{tab:corr}
	\end{subtable}
	\begin{subtable}{.5\linewidth}

		\begin{tabular}{c c c c c c c c c}
\toprule
		 & \textbf{ts1} & \textbf{ts2} & \textbf{ts3} & \textbf{ts4} & \textbf{ts5} & \textbf{ts6} & \textbf{ts7} & \textbf{ts8} \\
		\midrule
		ts1 & $\times$ & 0 & & & & & & \\
		ts2 & 0 & $\times$ & & & & & & \\
		ts3 & & & $\times$ & -1 & & & & \\
		ts4 & & & 1 & $\times$ & & & & \\
		ts5 & & & & & $\times$ & -1& -2& -3\\
		ts6 & & & & & 1 & $\times$ & -1 & -2 \\
		ts7 & & & & & 2 & 1 & $\times$ & -1 \\
		ts8 & & & & & 3 & 2 & 1 & $\times$ \\
		\bottomrule
		\end{tabular}
\caption{}
\label{tab:lags}
	\end{subtable}
\end{table}
\vspace{-12pt}

For the purpose of the experiments described in this paper, we used code from several sources: {\color{black}Matlab code that we developed for kernelised Geweke's measure and transfer entropy}, open access\linebreak  Matlab toolbox for Granger causality GCCA 
\citep{seth_matlab_2010,note3} and open access Matlab code provided by \linebreak Sohan Seth \citep{seth_assessing_2012,note4}.

To calculate Geweke's measure and kernelised Geweke's measure, we used the same code, with {a linear kernel} in the former case and {a Gaussian kernel in the latter. The} effect of regularisation on the (linear) Geweke's measure is negligible, {and} the results are comparable to the {ones obtained with} the GCCA code, with the main difference being more flexibility on the choice of lag ranges allowed by our code. Parameters for the ridge regression were either calculated with n-fold cross-validation ({see }Appendix~\ref{proced:crossval}) for the grid of regulariser values in the range of $[2^{-40}, \cdots, 2^{-26}]$ and kernel sizes in the range of $[2^{7}, \cdots, 2^{13}]$, or fixed at a {preset} level, with no noticeable impact on the result. Transfer entropy utilises {a} naive histogram to estimate distributions. {The} code for calculating HSNCIC and for performing \emph{p}-value tests incorporates {a} framework written by Seth \cite{seth_assessing_2012}. The framework has been altered to accommodate some new {functionalities}; the implementation of permutation tests has also been changed from rotation to actual permutation
\cite{note5} In the choice of parameters for the HSNCIC, we followed~\cite{seth_assessing_2012}, where the size of the kernel is set up as the median inter-sample distance and regularisation is set to $10^{-3}$.

{Our} goal was to uncover the causal structure without prior information and detect the lags at which causality occurred. This {was} performed by applying all three measures of causality with the following sets of lags: $\lbrace[1-10]{\rbrace, \lbrace}[1-20]\rbrace, \lbrace[1-5], [6-10], [11-15]\rbrace, \lbrace[1-3], [4-6], [7-9]\rbrace$; and finally, with all four measures to single lags $\lbrace 0,1,2,3,4\rbrace$. {Those ranges were used for linear and kernelised Geweke's measures and HSNCIC, but not for transfer entropy, for which only single lags are available with the current framework. Using five sets of lags allowed us to analyse the effects of using ranges of lags that are different from lags corresponding to the ``true'' dynamic of the variables.} {Table} \ref{table:cause:1lag} presents part of the results: \emph{p}-values for the four measures of interest for Lag 1. Below, we present the conclusions for each of the methods separately, with two Geweke's measures presented together.

Geweke's measures: Both Geweke's measures performed similarly, which was expected, as the data was simulated with linear dependencies. Causalities were correctly identified for all ranges of lags, for which the causal direction existed, including the biggest range [1--20]. For the shorter ranges \linebreak$\lbrace[1-5], [1-3]\rbrace$, as well as for the single lags $\lbrace 0,1,2,3\rbrace$, the two measures reported \emph{p}-values of zero for all of the existing causal directions. This means that the measures were able to detect precisely the lags at which causal directions existed, including Lag $0$, \textit{i.e.}, instantaneous coupling. {However,} with the number of permutations equal $200$ and at an acceptance level of $0.01$, the two measures {detected} only the required causalities, but would {fail to reject} some spurious causalities at a level of $0.05$.

Transfer entropy: By design, this measure can only analyse one lag at a time. It is also inherently slow, and for these two reasons, it will be inefficient when a wide range of lags needs to be considered. Furthermore, it cannot be used for instantaneous coupling{. In order to detect this, we applied the }mutual information method instead. For the lags $\lbrace 1,2,3\rbrace$, the transfer entropy reported zero \emph{p}-values for all the relevant causal directions. However, it {failed to reject} the spurious direction $1 \rightarrow 7$ with a \emph{p}-value of $0.01$. For lag $\lbrace 0\rbrace$, where mutual information has been applied, the instantaneous coupling $x_{1} \longleftrightarrow x_{2}$ {was} recognised correctly with a \emph{p}-value of zero.

HSNCIC: Due to slowness, HSNCIC is impractical for the largest ranges of lags. More importantly, HSNCIC performs unsatisfactorily for any of the ranges of lags that contained more than a single lag. This is deeply disappointing, as the design suggests HSNCIC should be able to handle both side information and higher dimensional variables. Even for a small range $[1--3]$, HSNCIC correctly recognised only the $x_{5} \rightarrow x_{8}$ causality. Nevertheless, it did recognise all of the causalities correctly when analysing one lag at a time, reporting \emph{p}-values of zero. {This suggests} that HSNCIC is unreliable for data that has more than one lag or more than two time series. HSNCIC is also not designed to detect instantaneous coupling.

\begin{table}[H]
\caption{\emph{p}-values for four measures {for Lag 1. From top left to bottom right: Geweke's measure (Gc), kernelised Geweke's measure (kG), transfer entropy (TE), HSNCIC (HS). All Lag 1 causalities were correctly retrieved by all methods.}}
\label{table:cause:1lag}
	\begin{subtable}{.5\linewidth}
	\centering
	\scriptsize
		\begin{tabular}{c|c c c c c c c c}
		Gc & ts1 & ts2 & ts3 & ts4 & ts5 & ts6 & ts7 & ts8 \\
		\hline
ts1 & 	$\times$ & 	0.97 & 	0.35 & 	0.26 & 	0.24 & 	0.68 & 	0.11 & 	0.23 \\
ts2 & 	0.42 & 	$\times$ & 	0.88 & 	0.52 & 	0.37 & 	0.69 & 	0.14 & 	0.46 \\
ts3 & 	0.26 & 	0.86 & 	$\times$ & 	0.75 & 	0.45 & 	0.19 & 	0.43 & 	0.72 \\
ts4 & 	0.14 & 	0.11 & 	\textbf{0} & 	$\times$ & 	0.24 & 	0.49 & 	0.41 & 	0.64 \\
ts5 & 	0.78 & 	0.94 & 	0.10 & 	0.02 & 	$\times$ & 	0.40 & 	0.96 & 	0.91 \\
ts6 & 	0.96 & 	0.31 & 	0.62 & 	0.22 & 	\textbf{0} & 	$\times$ & 	0.04 & 	1.00 \\
ts7 & 	0.74 & 	0.98 & 	0.10 & 	0.53 & 	0.35 & 	\textbf{0} & 	$\times$ & 	0.96 \\
ts8 & 	0.86 & 	0.70 & 	0.05 & 	0.63 & 	0.68 & 	0.87 & 	\textbf{0} & 	$\times$ \\
		\hline
		\end{tabular}
	\end{subtable}
	\begin{subtable}{.5\linewidth}
	\centering
	\scriptsize
		\begin{tabular}{c|c c c c c c c c}
		kG & ts1 & ts2 & ts3 & ts4 & ts5 & ts6 & ts7 & ts8 \\
		\hline
ts1 & 	$\times$ & 	0.92 & 	0.30 & 	0.25 & 	0.20 & 	0.74 & 	0.16 & 	0.16 \\
ts2 & 	0.50 & 	$\times$ & 	0.93 & 	0.54 & 	0.52 & 	0.71 & 	0.19 & 	0.46 \\
ts3 & 	0.29 & 	0.88 & 	$\times$ & 	0.68 & 	0.48 & 	0.11 & 	0.38 & 	0.62 \\
ts4 & 	0.12 & 	0.14 & 	\textbf{0} & 	$\times$ & 	0.22 & 	0.47 & 	0.41 & 	0.65 \\
ts5 & 	0.73 & 	0.93 & 	0.11 & 	0.04 & 	$\times$ & 	0.47 & 	0.99 & 	0.93 \\
ts6 & 	0.94 & 	0.38 & 	0.55 & 	0.18 & 	\textbf{0} & 	$\times$ & 	0.07 & 	0.99 \\
ts7 & 	0.81 & 	0.92 & 	0.04 & 	0.55 & 	0.36 & 	\textbf{0} & 	$\times$ & 	0.95 \\
ts8 & 	0.83 & 	0.67 & 	0.06 & 	0.63 & 	0.62 & 	0.86 & 	\textbf{0} & 	$\times$ \\

		\hline
		\end{tabular}
	\end{subtable}
 \newline
	\vspace*{0.25 cm}
	\newline
	\begin{subtable}{.5\linewidth}
	\centering
	\scriptsize

		\begin{tabular}{c|c c c c c c c c}
		TE & ts1 & ts2 & ts3 & ts4 & ts5 & ts6 & ts7 & ts8 \\
		\hline
ts1 & 	$\times$ & 	0.59 & 	0.53 & 	0.60 & 	0.34 & 	0.91 & 	0.38 & 	0.66 \\
ts2 & 	0.48 & 	$\times$ & 	0.86 & 	0.30 & 	0.87 & 	0.96 & 	0.49 & 	0.70 \\
ts3 & 	0.45 & 	0.17 & 	$\times$ & 	0.33 & 	0.34 & 	0.57 & 	0.81 & 	0.81 \\
ts4 & 	0.04 & 	0.31 & 	\textbf{0} & 	$\times$ & 	0.12 & 	0.09 & 	0.76 & 	0.08 \\
ts5 & 	0.21 & 	0.52 & 	0.86 & 	0.05 & 	$\times$ & 	0.68 & 	0.60 & 	0.30 \\
ts6 & 	0.53 & 	0.89 & 	0.65 & 	0.30 & 	\textbf{0} & 	$\times$ & 	0.77 & 	0.09 \\
ts7 & 	0.01 & 	0.42 & 	0.59 & 	0.37 & 	0.95 & 	\textbf{0} & 	$\times$ & 	0.77 \\
ts8 & 	0.85 & 	0.46 & 	0.07 & 	0.48 & 	0.85 & 	0.13 & 	\textbf{0} & 	$\times$ \\
		\hline
		\end{tabular}
	\end{subtable}
	\begin{subtable}{.5\linewidth}
	\centering
	\scriptsize

		\begin{tabular}{c|c c c c c c c c}
		HS & ts1 & ts2 & ts3 & ts4 & ts5 & ts6 & ts7 & ts8 \\
		\hline
ts1 & 	$\times$ & 	1.00 & 	0.81 & 	0.35 & 	0.19 & 	0.48 & 	0.71 & 	0.82 \\
ts2 & 	1.00 & 	$\times$ & 	0.80 & 	0.61 & 	0.85 & 	0.34 & 	0.02 & 	0.72 \\
ts3 & 	0.90 & 	0.95 & 	$\times$ & 	0.18 & 	0.59 & 	0.47 & 	0.21 & 	0.19 \\
ts4 & 	0.90 & 	0.29 & 	\textbf{0} & 	$\times$ & 	0.31 & 	0.81 & 	0.26 & 	0.31 \\
ts5 & 	0.75 & 	0.59 & 	0.77 & 	0.14 & 	$\times$ & 	0.71 & 	0.85 & 	0.46 \\
ts6 & 	0.64 & 	0.88 & 	0.75 & 	0.79 & 	\textbf{0} & 	$\times$ & 	0.71 & 	0.79 \\
ts7 & 	0.38 & 	0.13 & 	0.75 & 	0.24 & 	0.75 & 	\textbf{0} & 	$\times$ & 	0.60 \\
ts8 & 	0.90 & 	0.55 & 	0.46 & 	0.73 & 	0.78 & 	0.78 & 	\textbf{0} & 	$\times$ \\
		\hline
		\end{tabular}
	\end{subtable}
\end{table}

From this experiment, we conclude that Geweke's measures with linear and Gaussian kernels provide {the best performance, are not vulnerable to lag misspecification and seem the most practical.} The other two measures, transfer entropy and HSNCIC, provide good performance when analysing one lag at a time.
{In Section \ref{amblard_example} we show the results of one of the tests from \cite{amblard_kernelizing_2012}, which investigates the ability to distinguish between direct and non-direct causality in data where both linear and non-linear dependence have been introduced. We refer to \cite{seth_assessing_2012} for the results of a wide range of tests applied to linear Granger causality and HSNCIC. We tested all four methods and managed to reproduced the results from \cite{seth_assessing_2012} to a large degree; however, we used smaller number of permutations and realisations, and we obtained somewhat lower acceptance rates for true causal directions, particularly for HSNCIC.} From all of those {tests, we conclude} that linear causality can be detected by all measures in most cases, with the exception of HSNCIC when more lags or dimensions are present. Granger causality can detect some nonlinear causalities, especially if they can be approximated by linear functions. Transfer entropy will flag more spurious causalities in the case where causal effects exist for different lags. There is no maximum dimensionality that HSNCIC can accept; in some experiments, this measure performed well for three and four dimensional problems; in others, three dimensions proved to be too many.

Possibly the most important conclusion is that parameter selection turned out to be critical for kernelised Geweke's measure. For some tests, like the simulated eight time series data described earlier, the size of the kernel did not play an important role, but in some cases, {the} size of the kernel was crucial in allowing the detection of causality. However, there was no kernel size that {worked} for all of the types of the data.

{
\subsubsection{Nonlinear Multivariate Example}
\label{amblard_example}

Our second example follows one presented by Amblard \citep{amblard_kernelizing_2012} and involves a system with both linear and non-linear causality. Apart from presenting the benefits of generalising Granger causality, this example demonstrates the potential effect of considering side information on distinguishing direct and indirect cause.
The true dynamic of the time series is as follows:
\begin{equation}
\begin{split}
\begin{cases}
x_{t} = a x_{t-1} + \epsilon_{x,t}\\
y_{t} = b y_{t-1} + d x_{t-1}^{2} + \epsilon_{y,t}\\
z_{t} = c z_{t-1} + e y_{t-1} + \epsilon_{z,t}
\end{cases}
\end{split}
\end{equation}
where the parameters were chosen in the following way: $a = 0.2, b = 0.5, c = 0.8, d = 0.8, e = 0.7$, the variables $\epsilon_{x,t}, \epsilon_{y,t}, \epsilon_{z,t}$ are i.i.d. Gaussian with zero mean and unit variance. From the setup, we know that we have the following causal chain $x \rightarrow y \rightarrow z$ (with the nonlinear effect of $x$ on $y$), and therefore, there is an indirect causality $x \rightarrow z$. We calculate kernelised Geweke measures $G_{x\rightarrow z}$ and $G_{x\rightarrow z \mid y}$ to assess the~causality.

We repeat the experiment 500 times, each time generating a time series of length 500. We choose an embedding of two, \textit{i.e.}, we consider the lag range $[1--2]$. To evaluate the effect of using kernelised rather than linear Granger causality, we run each experiment for the Gaussian kernel and for linear kernel $k(x,y) = x^{T}y$. Using the linear kernel is nearly equivalent to using the linear Geweke measures.
We obtain a set of 500 measurements for $G_{x\rightarrow z}$ and $G_{x\rightarrow z \mid y}$, each run with a Gaussian and with a linear kernel. The results are shown in Figures \ref{fig:link:Amblard}--\ref{fig:HSNCIC:Amblard}. As expected, $G_{x\rightarrow z \mid y}$ does not detect any causality, regardless of the kernel chosen. When no side information is taken into consideration, we should see the indirect causality $x \rightarrow z$ being picked up; however, this is the case only for kernelised Geweke with the Gaussian kernel and for HSNCIC. As the dependence was nonlinear, the linear Geweke's measure did not \linebreak detect it.

\begin{figure}[H]
 \centering
 \includegraphics[height = 7.5cm]{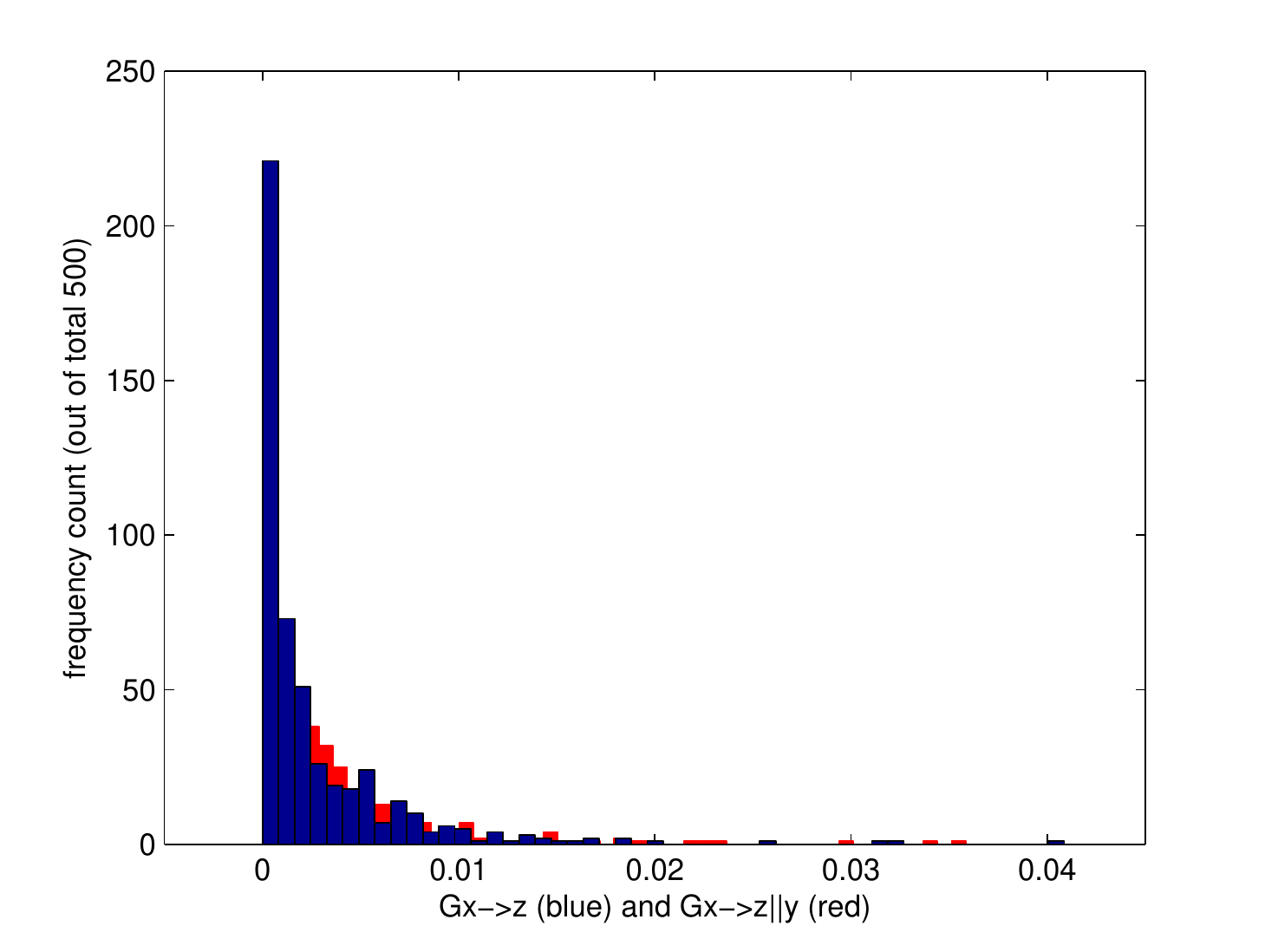}
 \caption{Histogram of the measurements $G_{x\rightarrow z}$ (red face), $G_{x\rightarrow z \mid y}$ (blue face), calculated with the kernelised Geweke's using the linear kernel (\textit{i.e.}, the equivalent of\protect\linebreak  Granger causality).}
 \label{fig:link:Amblard}
\end{figure}

\begin{figure}[H]
 \centering
 \includegraphics[height =8cm]{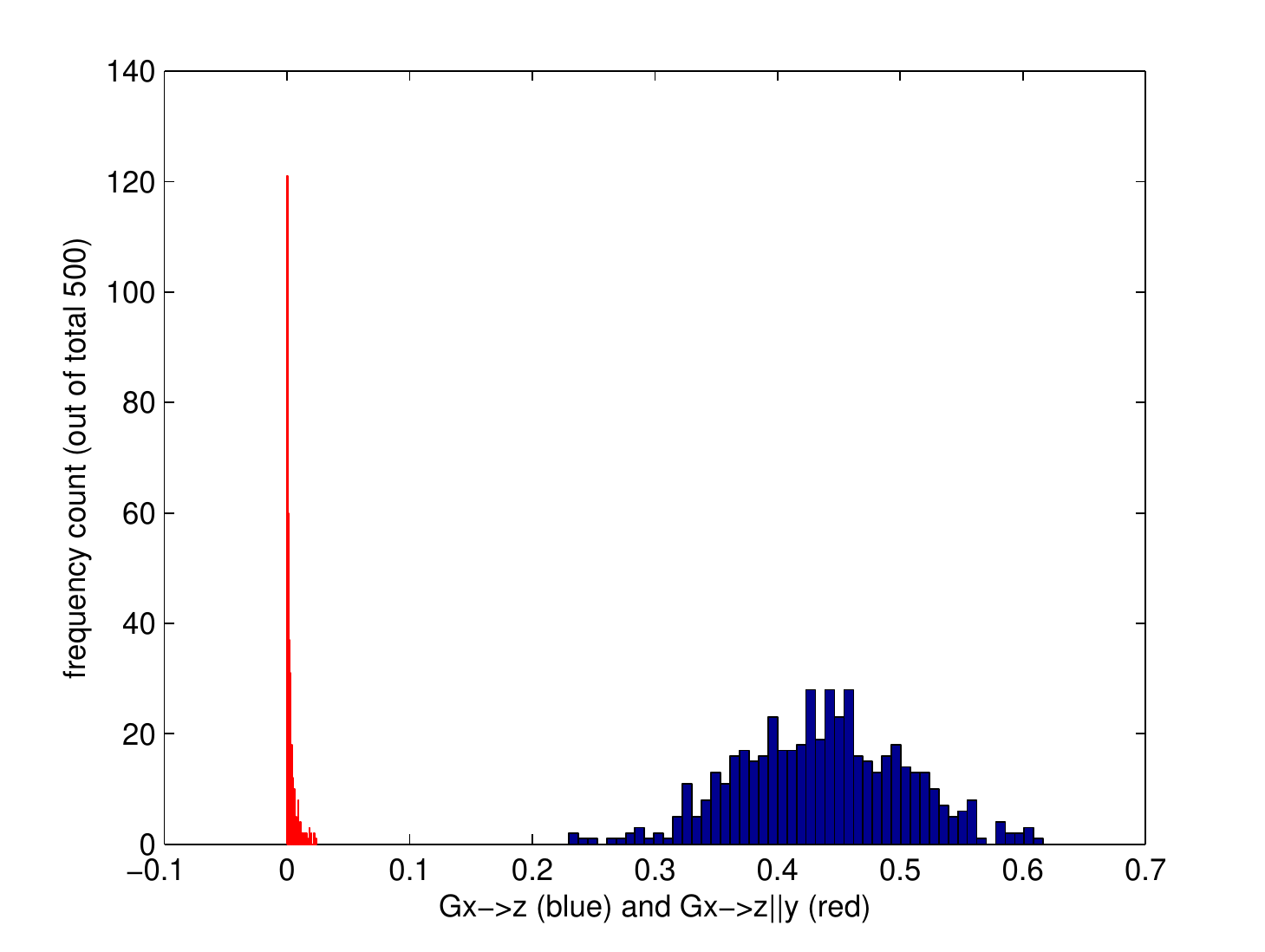}
 \caption{Histogram of the measurements $G_{x\rightarrow z}$ (red face), $G_{x\rightarrow z \mid y}$ (blue face), calculated with the kernelised Geweke's using the Gaussian kernel.}
 \label{fig:Gaussk:Amblard}
\end{figure}

\begin{figure}[H]
 \centering
 \includegraphics[height = 8cm]{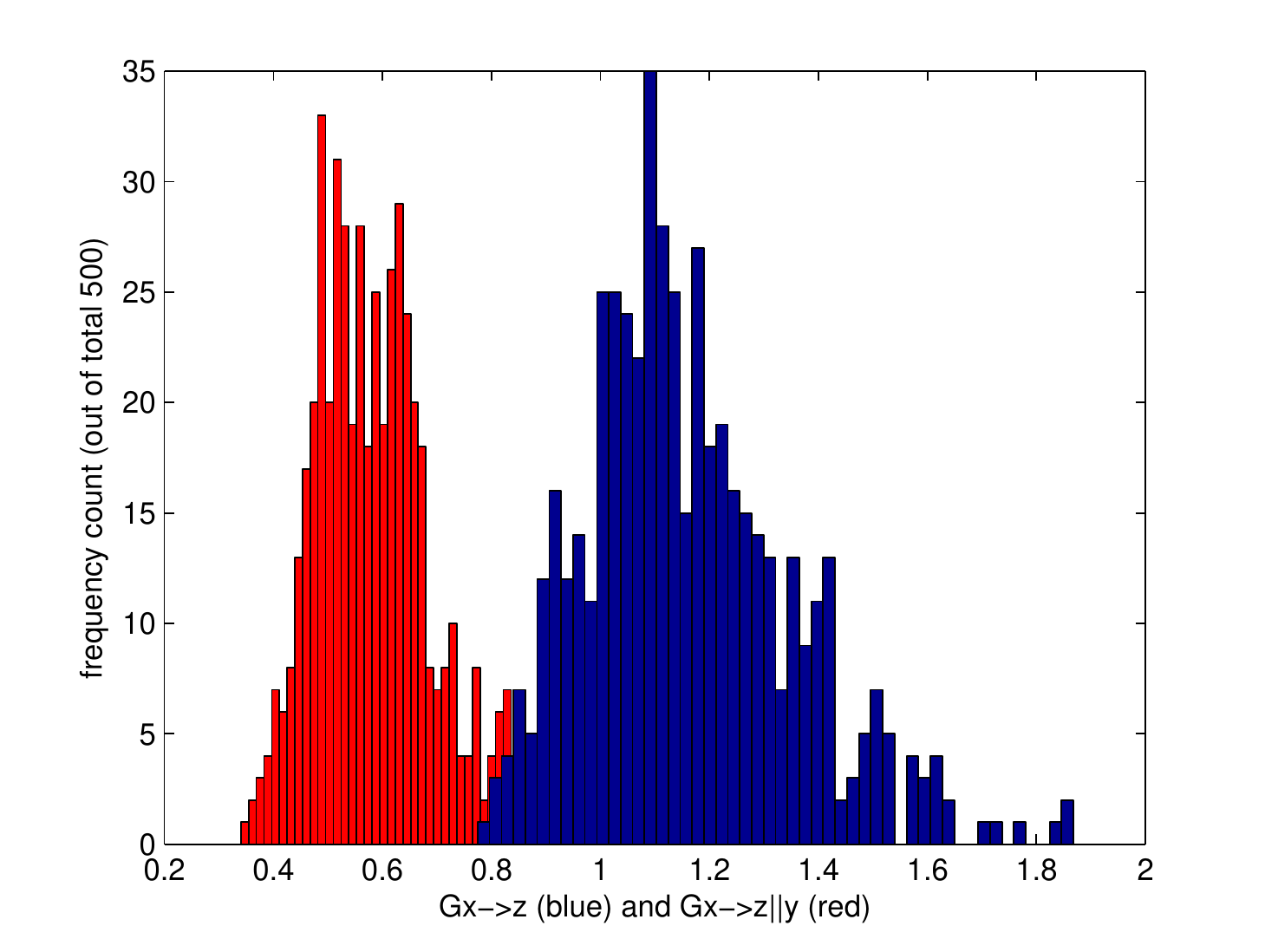}
 \caption{Histogram of the measurements $G_{x\rightarrow z}$ (red face), $G_{x\rightarrow z \mid y}$ (blue face), calculated with the Hilbert--Schmidt Normalised Conditional Independence Criterion (HSNCIC).}
 \label{fig:HSNCIC:Amblard}
\end{figure}

Transfer entropy, as defined in this paper, does not allow side information, and therefore, the result we achieve is a distribution that appears significantly different from zero (Figure \ref{fig:TE}).

\begin{figure}[H]
 \centering
 \includegraphics[height = 8cm]{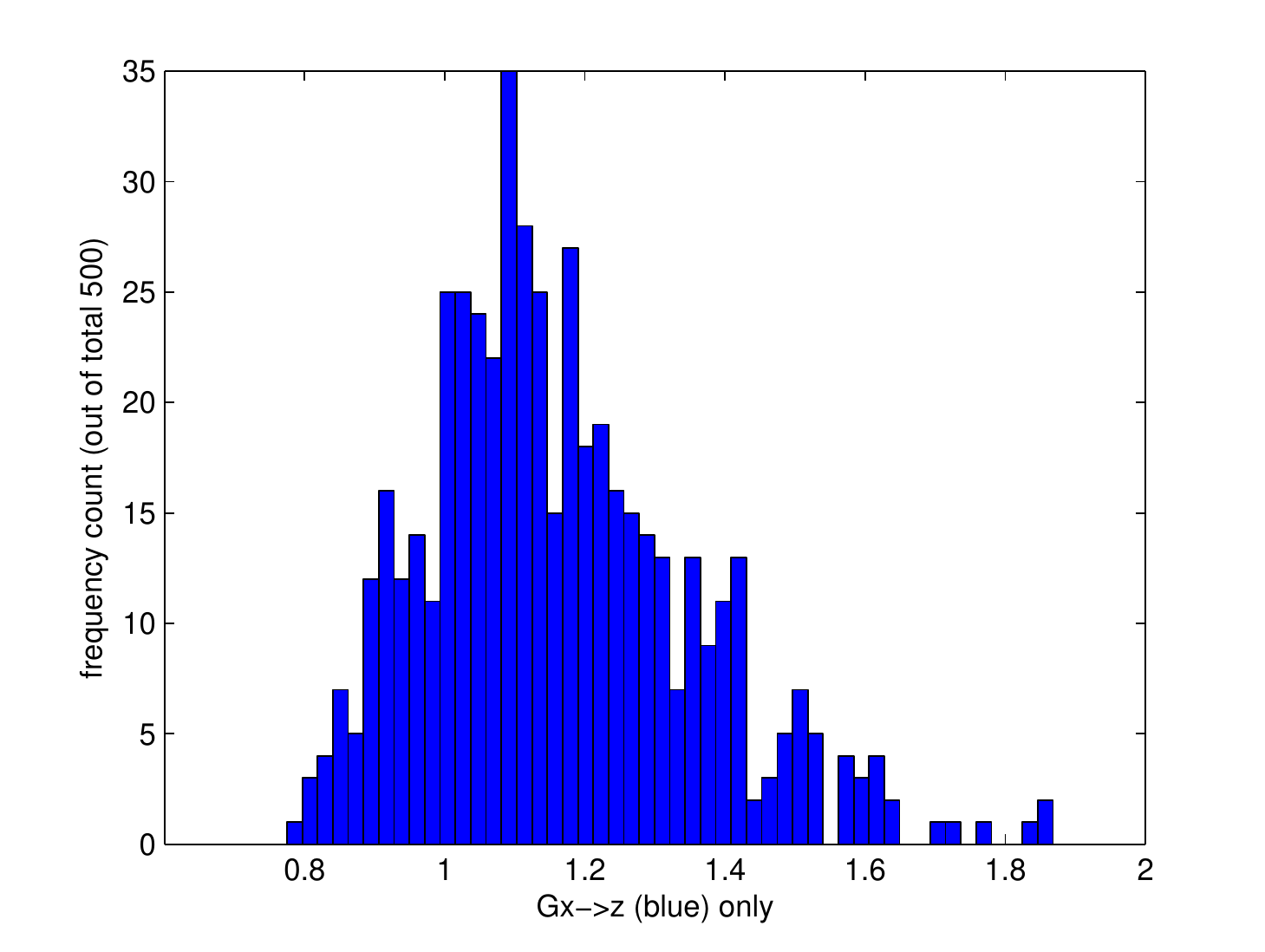}
 \caption{Histogram of the measurements $G_{x\rightarrow z}$ (red face), $G_{x\rightarrow z \mid y}$ (blue face), calculated with the transfer entropy.}
 \label{fig:TE}
\end{figure}

} 

\section{Applications}
\label{applications}

Granger causality {was} introduced as an {econometrics} concept, and for many years, it was mainly used in economic applications.
After around 30 years of relatively little acknowledgement, the concept of causality started to gain significance in a number of scientific disciplines. Granger causality and its generalisations and alternative formulations became popular, particularly in the field of neuroscience, but also climatology and physiology \cite{hlavackova-schindler_causality_2007,amblard_directed_2011,chavez_statistical_2003,knuth_revealing_2005, gourevitch_evaluating_2007,vicente_transfer_2011}. The methodology {was} successfully applied in those fields, particularly in neuroscience, due to the characteristics of the data common in those fields and the fact that the assumptions of Gaussian distribution and/or linear dependence are often reasonable \citep{bressler_wiener-granger_2011}. This is generally not the case for financial time series.

\subsection{Applications to Finance and {Economics}}

In finance and {economics}, there are many tools devoted to modelling dependence, mostly for symmetrical dependence, such as correlation/covariance, cointegration, copula {and, to a lesser} degree, mutual information \cite{alexander_cointegration_1994,cont_long_2005,patton_copulabased_2009, durante_copulas_2013}. However, in various situations where we would like to reduce the dimensionality of a problem (e.g., choose a subset of instruments to invest in, choose a subset of variable for a factor model, \emph{etc}.), knowledge of the causality structure can help in choosing the most relevant dimensions. {Furthermore}, forecasting using the causal time series (or Bayesian priors in Bayesian models or parents in graphical models \citep{pearl_causality:_2000,barber_bayesian_2012}) helps to forecast ``future rather than the past''.

Financial data often {have} different characteristics than data most commonly analysed in biology, physics, \emph{etc}. In finance, the typical situation is that the researcher has only one long, multivariate time series at her disposal, while in biology, even though the experiments might be expensive, most likely, there will be a number of them, and usually, they can be reasonably assumed to be independent identically distributed ({i.i.d.}). The assumption of linear dependencies or Gaussian distributions, often argued to be {reasonable} in disciplines, such as neuroscience, are commonly thought to be invalid for financial time series. Furthermore, many researchers point out that stationarity usually does not apply to this kind of data. As causality methods in most cases assume stationarity, the relaxation of this requirement is clearly an important direction for future research. In the sections below, we describe the results of applying causal methods to two sets of financial data.

\subsubsection{Interest Rates and Inflation}

Interest rates and inflation have been investigated by economists for a long time. There is considerable research concerning {the} relationship between inflation and nominal or real interest rates for the same country or region, some utilising tools of Granger causality (for example, \cite{eichler_granger_2007}).

{In this experiment, we analyse} related values, namely the consumer price index for the United States (U.S. CPI) and the London Interbank Offered Rate (Libor) interest rate index. Libor is often used as a base rate (benchmark) by banks and other financial institutions, and it {is} an important economic indicator. It is not a monetary measure associated with any country, and it does not reflect any {institutional} mandate in contrast to, for example, when the Federal Reserve sets interest rates. Instead, it reflects some level of assessment of risk by the banks who set the rate. Therefore, we ask {whether} we detect that one of these two economic indicators causes the other one in a statistical sense?

We {ran} our analysis for monthly data from January 31, 1986, to October 31, 2013, obtained from Thomson Reuters. {The} implementation and {parameter values} used for this analysis {were} similar {to those in} the simulated example ({Section} \ref{testing:simdata}). We used kernelised Geweke's measure with linear and Gaussian kernels. Parameters for the ridge regression were at a {preset} level in the range of $[2^{7}, \cdots, 2^{13}]$ or as \linebreak a median.

\begin{figure}[H]
\centering
 \includegraphics[width=14cm, height = 8cm]{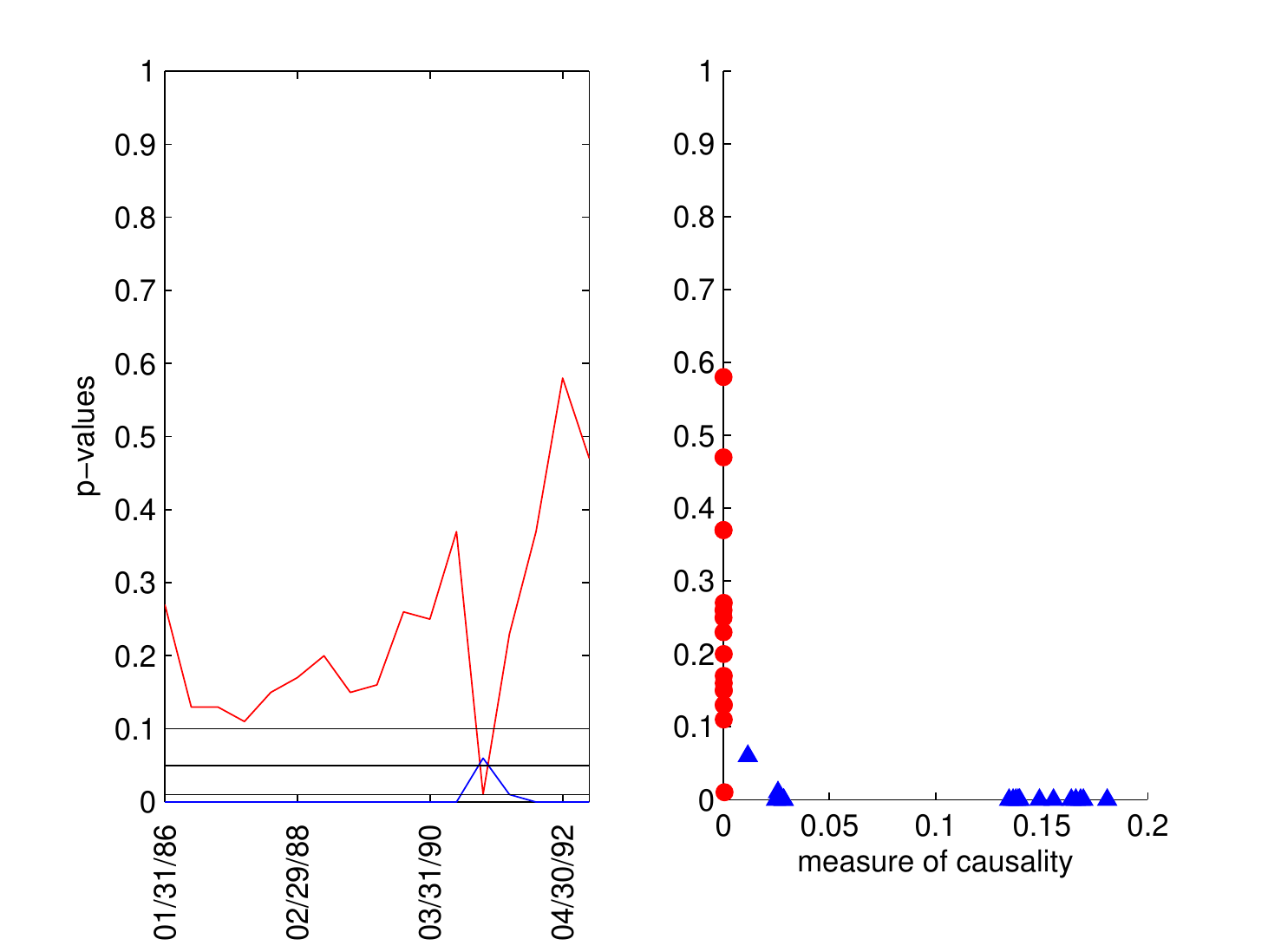}
 \caption{Kernelised Geweke's measure of causality. The left chart shows sets of \textit{p}-values for the hypothesis that inflation statistically causes Libor (blue line) or the other way round {(red line)}, when a model with one lag is considered. The right chart shows the scatter plot of \emph{p}-values and the value of the causality measure.}
 \label{fig:ii1}
\end{figure}

\begin{figure}[H]
\centering
 \includegraphics[width=14cm, height = 8cm]{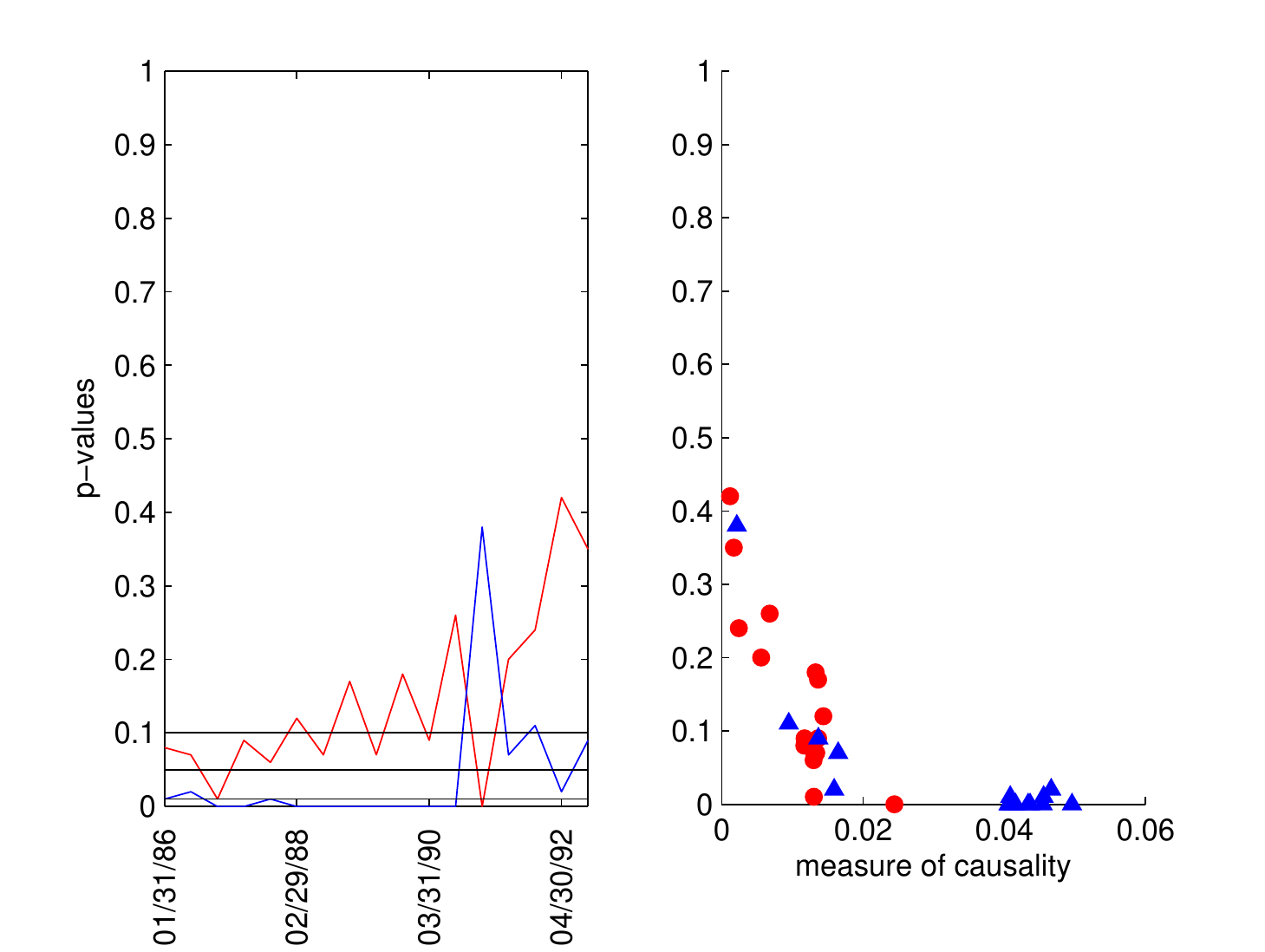}
 \caption{Linear Geweke's measure of causality. (\textbf{Left}) Sets of \textit{p}-values for the hypothesis of statistical causality in the direction U.S. consumer price index $\rightarrow$ one-month Libor (blue line) or the other way round (red line), when a model with a linear kernel and Lag 1 is considered. (\textbf{Right}) Scatter plot of \emph{p}-value and value of the causality measure.}
 \label{fig:ii3}
\end{figure}

\begin{figure}[H]
\centering
 \includegraphics[width=14cm, height = 9cm]{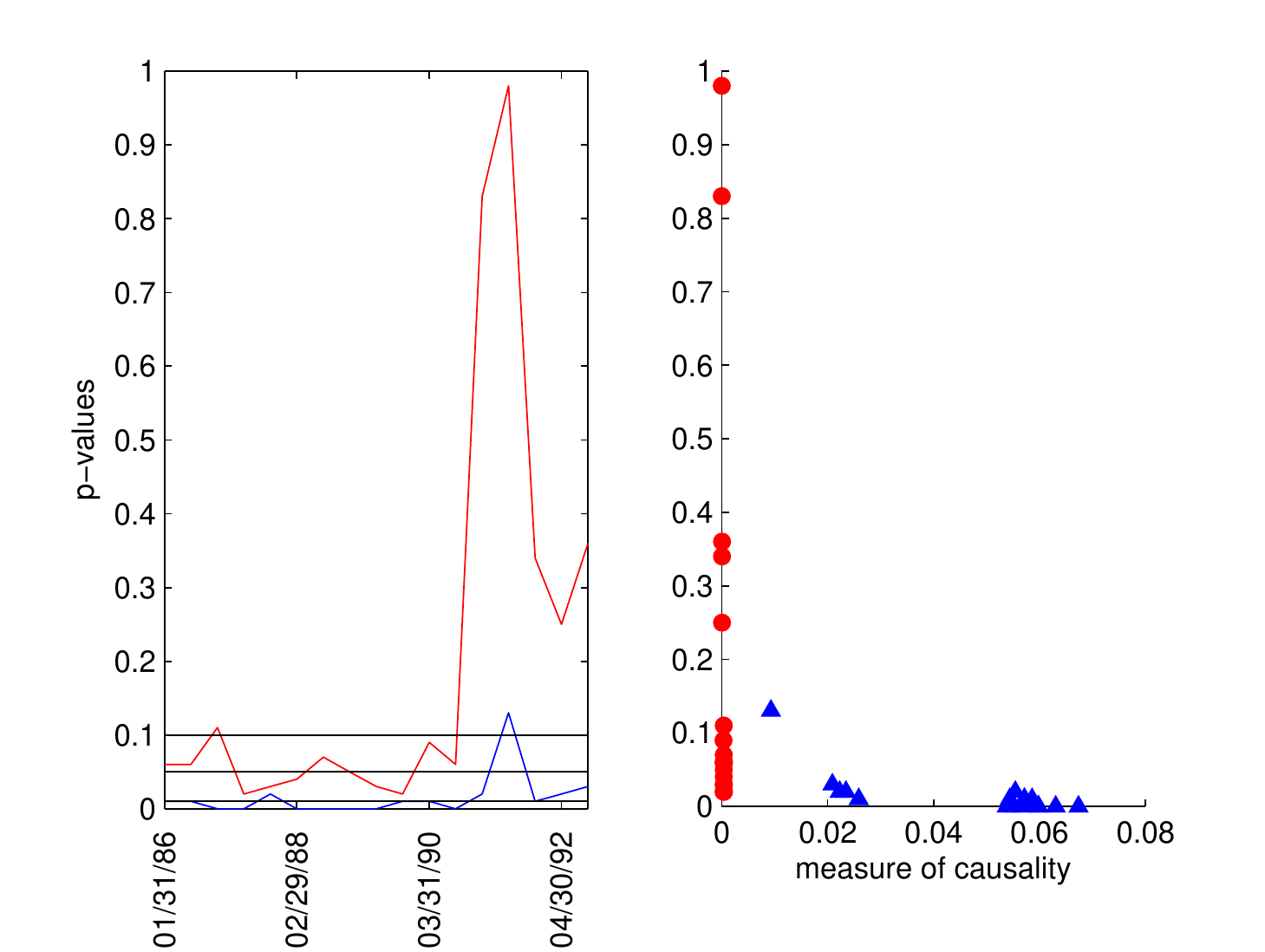}
 \caption{Kernelised Geweke's measure of causality. (\textbf{Left}) Sets of \textit{p}-values for the hypothesis of statistical causality in the direction U.S. CPI $\rightarrow$ one-month Libor (blue line) or the other way round (red line), when {the} model with the Gaussian kernel and Lag 2 is considered; (\textbf{Right}) Scatter plot of the \emph{p}-value and the value of the causality measure.}
 \label{fig:ii2}
\end{figure}

\begin{figure}[H]
\centering
 \includegraphics[width=14cm, height = 9cm]{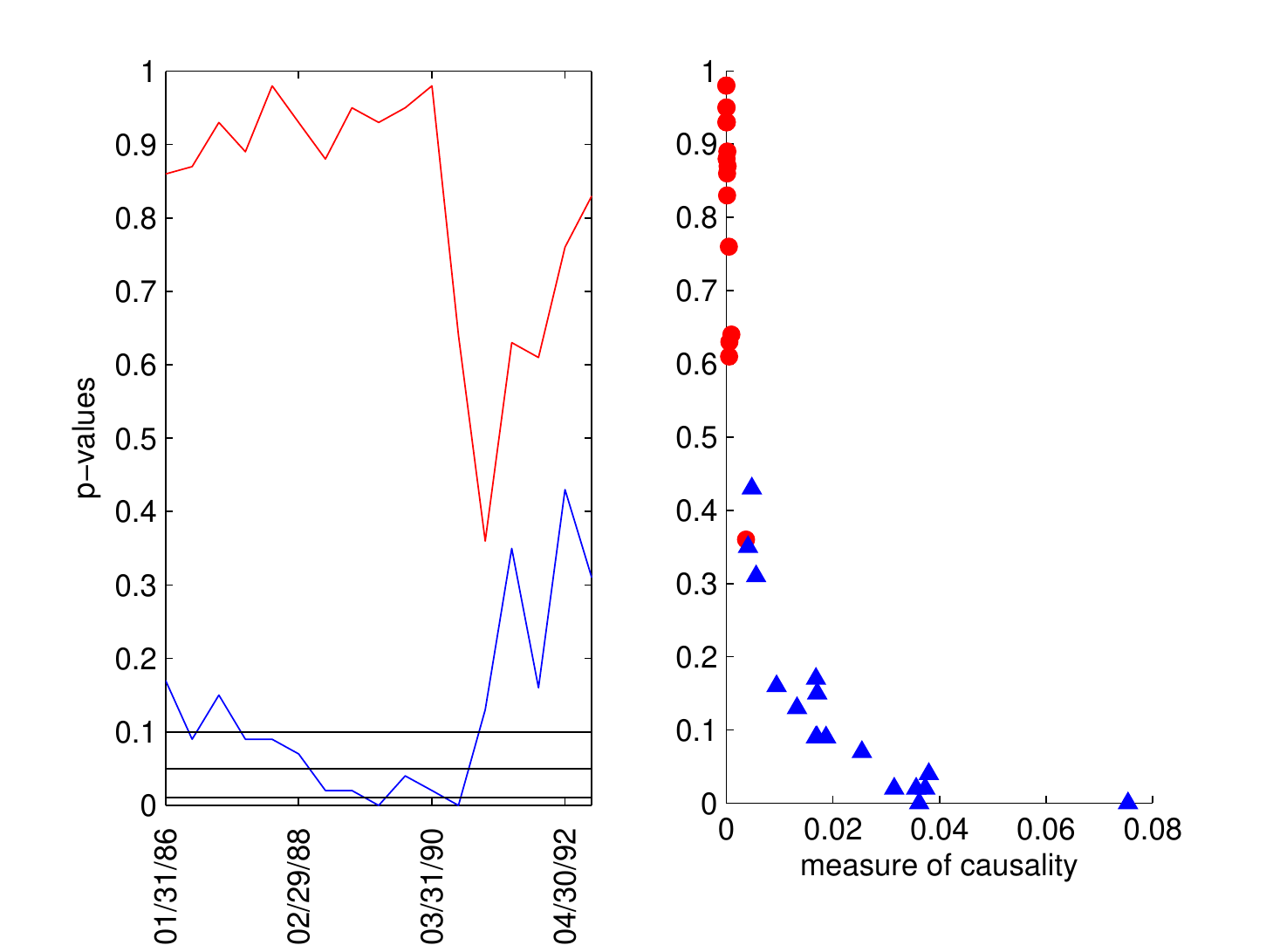}
 \caption{Linear Geweke's measure of causality. (\textbf{Left}) Sets of \textit{p}-values for the hypothesis of statistical causality in the direction U.S. CPI $\rightarrow$ one-month Libor (blue line) or the other way round (red line), when model with a linear kernel and Lag 7 is considered; (\textbf{Right}) Scatter plot of the \emph{p}-value and the value of the causality measure.}
 \label{fig:ii4}
\end{figure}

We investigated {time-windows of size} 25, 50, 100 and 250. The most statistically significant and interpretable results {were} observed for the longer windows (250 points), where Geweke's measure and kernelised Geweke's measure show a clear indication of the direction U.S. CPI $\rightarrow$ Libor. For shorter windows of time, significant \emph{p}-values were obtained considerably less often, but the results were consistent with the results for the longer time window. The dependence for the 250 day window {were seen} most strongly for Lag 1 (\textit{i.e.}, one month) and less strongly for Lags 2, 7, 8, 9, but there is no clear direction for the interim lags. In {Figures} \ref{fig:ii1}--\ref{fig:ii4}, we report \emph{p}-values for the assessment of causality for Lags 1, 2 and 7 alongside the scatter plot showing \emph{p}-values and the values of Geweke's measure. All of the charts have been scaled to show \emph{p}-values in the same range [0,1]. We can clearly see the general trend that the higher {the} values of causality, the lower their corresponding \emph{p}-values.

In {Figure} \ref{fig:ii1}, we observe that the U.S. CPI time series lagged by one month causes one-month Libor in a statistical sense, when assessed with kernelised Geweke's measure with Gaussian kernel. The \emph{p}-values for the hypothesis of causality in this direction allow {us} to accept {(not reject)} this hypothesis at a significance level of $0.01$ in most cases, with the \emph{p}-values nearly zero most of the time. We can also observe that several of the causality measurements are as high as $0.2$, which can be translated to {an improvement of roughly $0.18$} in the explanatory power of the model \cite{note6}.
Applying the linear kernel ({Figure} \ref{fig:ii3}) resulted in somewhat similar patterns of measures of causality and \emph{p}-values, but the two directions were less separated. Interest rates causing Libor still have \emph{p}-values at zero most of the time, but the other direction has \emph{p}-values that fall below the 0.1 level for several consecutive windows at the beginning, with the improvement in the explanatory power of the model at a maximum 0.07 level; our interpretation is that the causality is nonlinear.

The results for the second lag, given in {Figure} \ref{fig:ii2}, are no longer as clear as for Lag 1 in {Figure} \ref{fig:ii1} (Gaussian kernel in both cases). The hypothesis of inflation causing interest rates still has \emph{p}-values close to zero most of the time, but the \emph{p}-values for the other direction are also small. This time, the values of causality are lower and reach up to {just} below $0.08$. Using a linear kernel, we obtain less clear results, and our interpretation is that the causal direction CPI $\rightarrow$ Libor is stronger, but there might be some feedback, as well.

{Figure \ref{fig:ii4} presents the} results of using {a} linear kernel, which {shows} a much better separation of the two directions, applied to the model with Lag 7. Very similar results can be seen for models with Lags 8 and 9. \linebreak  There is no {obvious} reason why {the} linear kernel performed much better than {the} Gaussian kernel {for these large lags}. We offer the interpretation that no nonlinear causality was strong enough and consistent enough and that this {was} further obscured by using a nonlinear kernel. The conclusion here is that model selection is an important aspect of detecting causality and needs {further} research.

\begin{figure}[H]
\centering
 \includegraphics[width=15cm, height = 9cm]{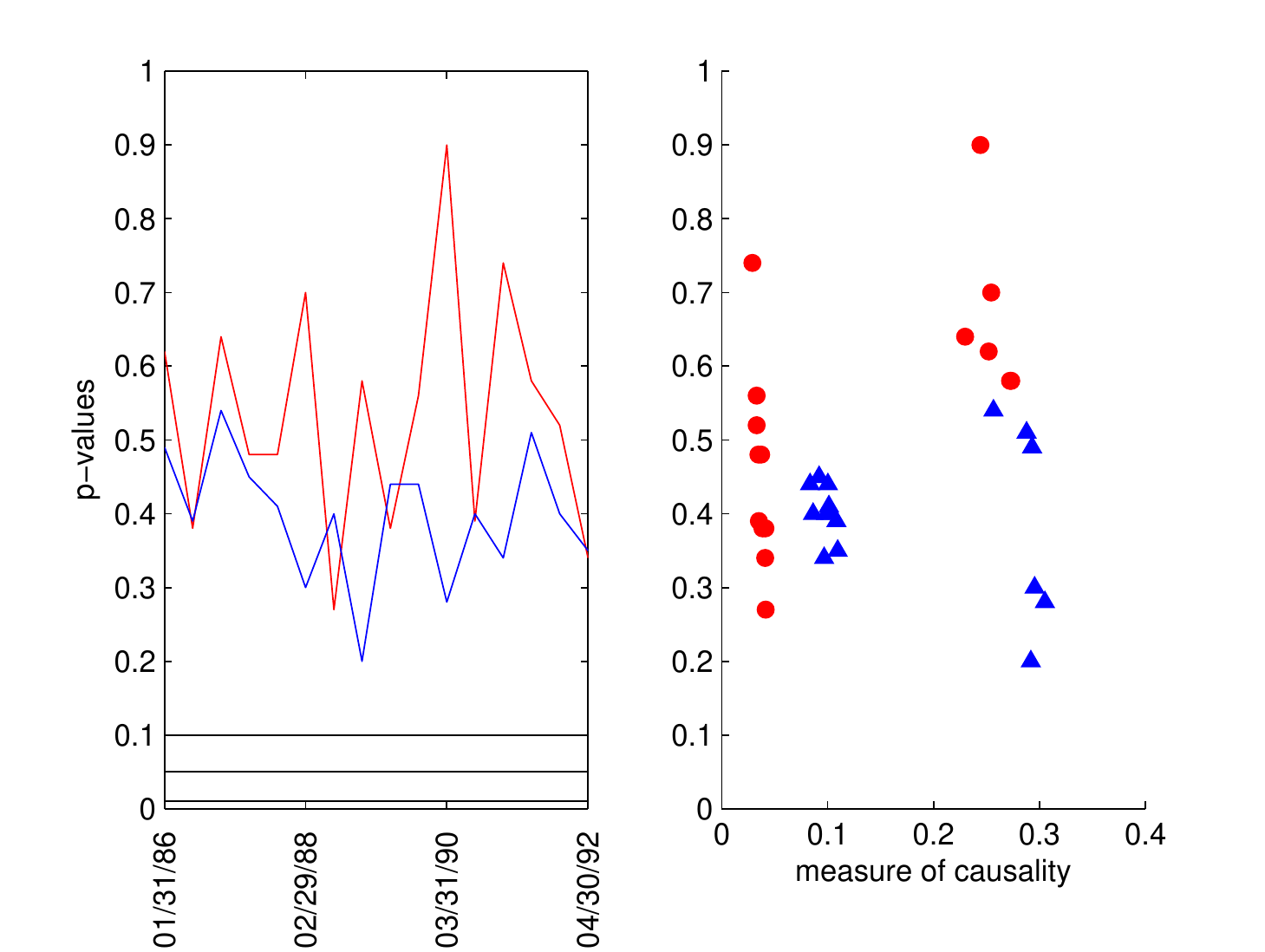}
 \caption{Transfer entropy. (\textbf{Left}) sets of \textit{p}-values for the hypothesis of statistical causality in the direction U.S. CPI $\rightarrow$ one-month Libor (blue line) or the other way round (red line), when Lag 1 is considered; (\textbf{Right}) Scatter plot of the \emph{p}-value and the value of the causality~measure.}
 \label{fig:teLlag1}
\end{figure}

\begin{figure}[H]
\centering
 \includegraphics[width=15cm, height = 9cm]{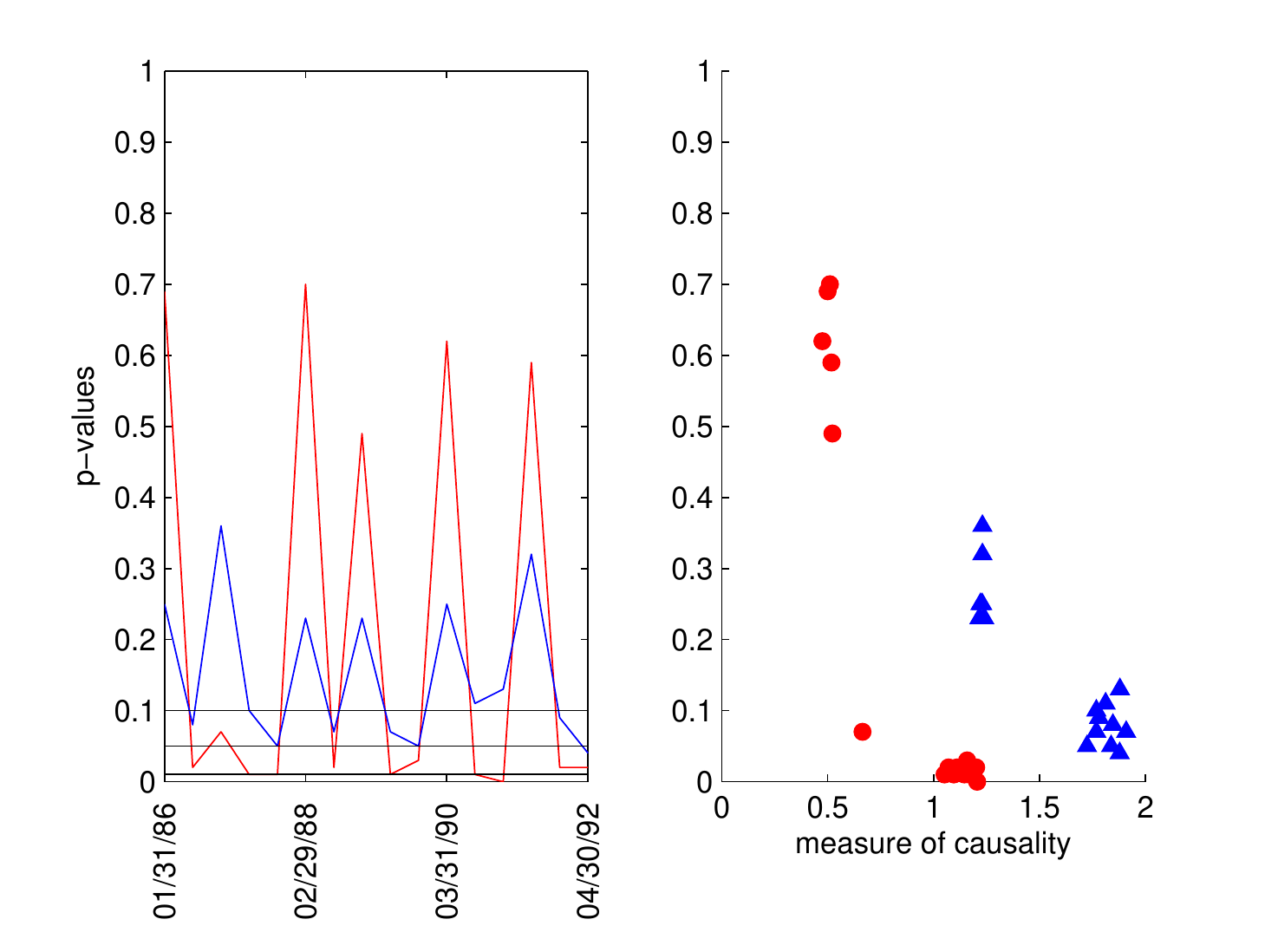}
 \caption{HSNCIC. (\textbf{Left}) sets of \textit{p}-values for the hypothesis of statistical causality in the direction U.S. CPI $\rightarrow$ one-month Libor (blue line) or the other way round (red line), when Lag 1 is considered; (\textbf{Right}) Scatter plot of the \emph{p}-value and the value of the\protect\linebreak causality measure.}
 \label{fig:hsLlag1}
\end{figure}

In our analysis, we {did not obtain significant results for transfer entropy or HSNCIC}. The results for Lag 1 for transfer entropy and HSNCIC are shown in {Figures} \ref{fig:teLlag1} and \ref{fig:hsLlag1}, respectively. For Lag 1, there was a significant statistical causality in the direction U.S. CPI $\rightarrow$ one-month Libor supported by both Geweke's measures. This is barely seen for transfer entropy and HSNCIC. \emph{p}-values for transfer entropy are at {a} level that only slightly departs from a random effect, and for HSNCIC, they are often significant; however, the two directions are not well separated. The results for higher lags were often even more difficult to interpret. {We must stress that the different implementation of transfer entropy and parameter choice for HSNCIC might result in better performance (please refer to Sections \ref{theo_diff} and \ref{mod:selection}).}

\subsubsection{Equity \emph{versus} Carry Trade Currency Pairs}

We analysed six exchange rates (AUDJPY, CADJPY, NZDJPY, AUDCHF, CADCHF, NZDCHF and the index S$\&$P) and investigated any patterns of the type ``leader-follower''. {Our} expectation was that S$\&$P should be leading. We used daily data for the period July 18, 2008--October 18, 2013, from Thomson Reuters. We {studied} the pairwise dependence between the currencies and S$\&$P, and we {also analysed} the results of adding the Chicago Board Options Exchange Market Volatility Index (VIX) as side information. In all of the cases, we {used} logarithmic returns.

%

\begin{figure}[H]
\centering
 \includegraphics[width=18cm, height = 12cm]{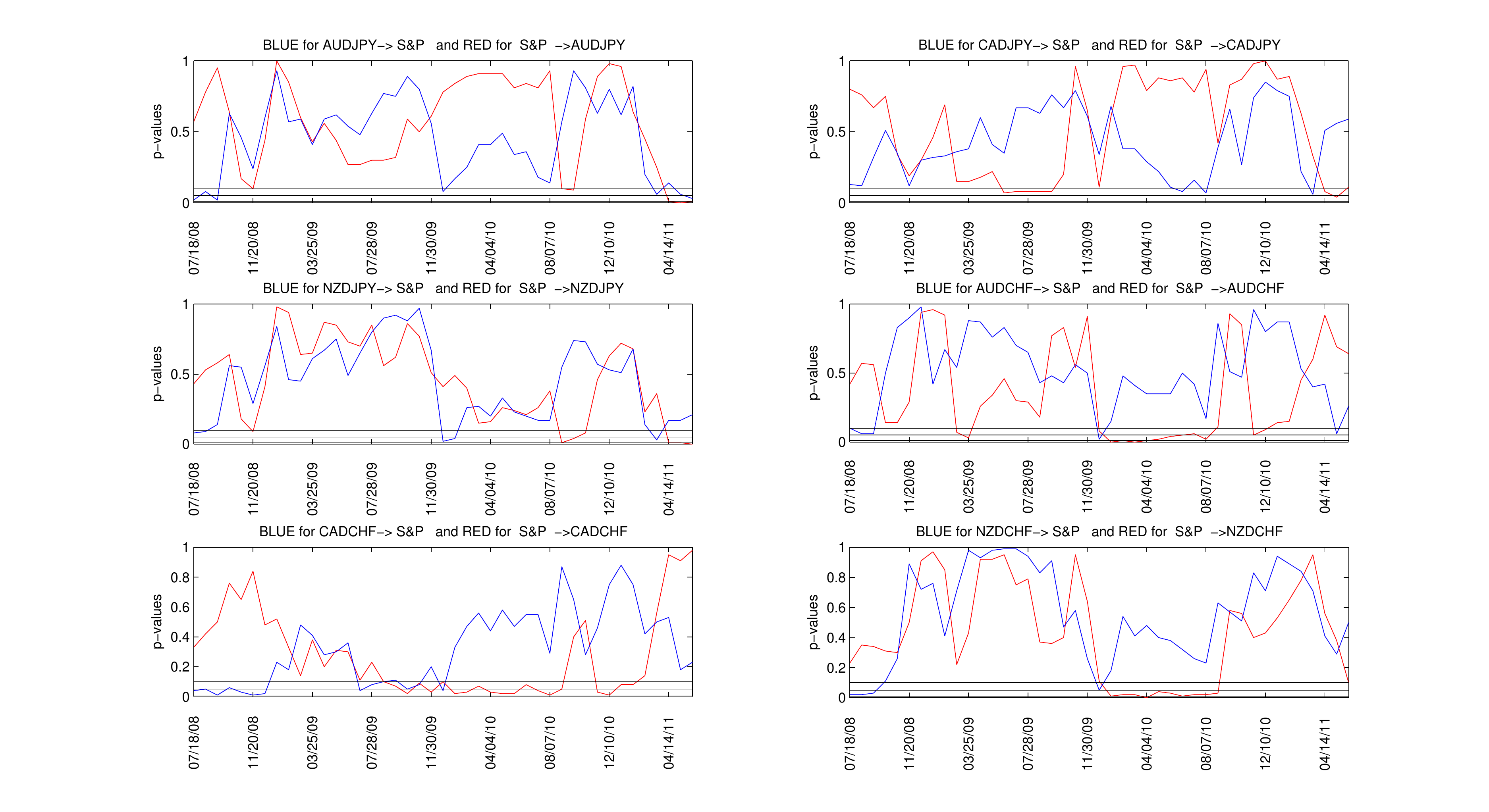}
 \caption{Sets of \textit{p}-values for the hypothesis that an exchange rate causes the equity index, S$\&$P (blue), or the other way round (red).}
 \label{fig:fxeq_1}
\end{figure}

Figure \ref{fig:fxeq_1} presents the results of applying kernelised Geweke's measure with a Gaussian kernel. The plots show series of \emph{p}-values for a moving window of a length of 250 data points (days), {moving each window by 25 points}. Unlike in the previous case of interest rates and inflation, there is little actual difference between {the} linear and Gaussian kernel {methods. However, in a few cases, employing a} Gaussian kernel results in better separation of the two directions, especially CADCHF $\rightarrow$ S$\&$P and S$\&$P $\rightarrow$ CADCHF given VIX.

{Excepting} CADCHF{,} all currency pairs exhibit similar behaviour when analysed for the causal effect on {the} S$\&$P. This behaviour consists of a small number of windows for which a causal relationship is significant at a \emph{p}-value below 0.1, but that {does not} persist. CADCHF is the only currency with a consistently significant causal effect on S$\&$P, which is indicated for periods starting in $2008$ and $2009$. As for the other direction, for AUDCHF, CADCHF and NZDCHF, there are periods where S$\&$P has a significant effect on them as measured by \emph{p}-values.

We obtained all of the main ``regimes'': Periods when either one of the exchange rates or S$\&$P had more explanatory power (\emph{p}-values for one direction were much lower than for the other) and periods when both exhibited low or both exhibited high \emph{p}-values. \emph{p}-values close to one {did not} necessarily mean {purely a} lack of causality: in such cases, the random permutations of the time series tested for causality at a specific lag appear to have higher explanatory power than the time series at this lag itself. There are a few possible explanations related to the data, the measures and to the nature of the permutation test itself. We {observed} on the simulated data that when no causality is present, autocorrelation introduces biases to the permutation test: higher \emph{p}-values than we would expect from a randomised sample, but also the higher likelihood of interpreting correlation as causality. Furthermore, both of these biases can result from assuming {a model with a lag different from that of the data.} Correspondingly, if the data has been simulated with instantaneous coupling and no causality, this again can result in high \emph{p}-values. Out of all four methods, transfer entropy appeared to be most prone to these biases.

\begin{figure}[H]
\centering
 \includegraphics[width=1\linewidth]{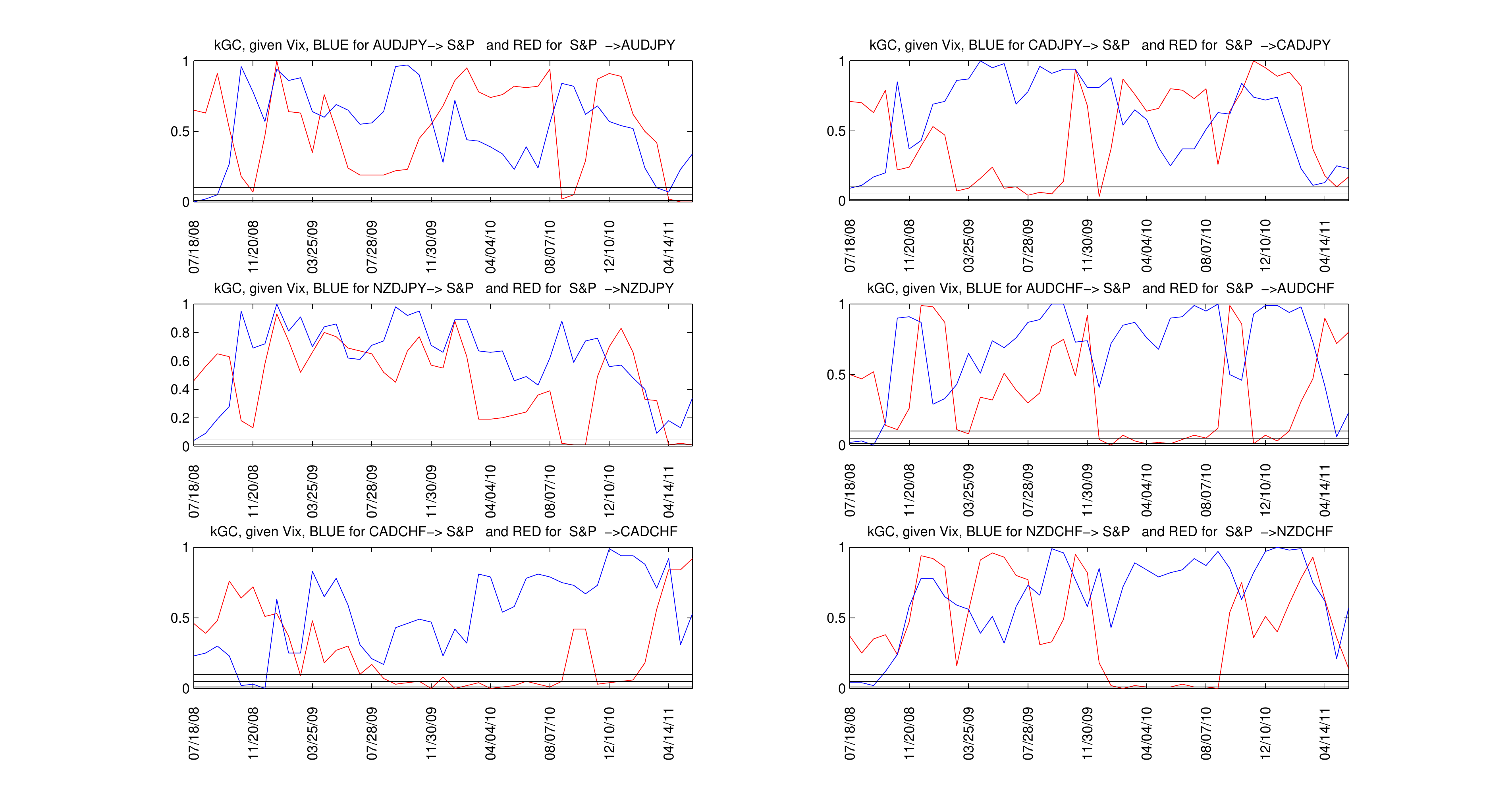}
 \caption{Sets of \textit{p}-values for the hypothesis that an exchange rate causes the equity index, S$\&$P, given the Volatility Index (VIX) as side information (blue) or the other way round (red).}
 \label{fig:fxeqVIX}
\end{figure}

Figure \ref{fig:fxeqVIX} shows similar information as in {Figure} \ref{fig:fxeq_1}, but taking into consideration VIX as side information. The rationale is that the causal effect of S$\&$P on the carry trade currencies is likely to be connected to the level of perceived market risk. However, the charts {do not} show the disappearance of a causal effect after including VIX. While the patterns {do not} change considerably, we observe that exchange rates have lost most of their explanatory power for S$\&$P, with the biggest differences for CADCHF. There is little difference for the \emph{p}-values for the other direction; hence, the distinction between the two directions became more significant.

%


\section{Discussion and Perspectives}
\label{discussion}

While questions about causal relations are asked often in science in general, the appropriate methods of quantifying causality for different contexts are not well developed. Firstly, often, answers are formulated with methods not intended specifically for this purpose. There are fields of science, for example nutritional epidemiology, where causation is commonly inferred from correlation \cite{note7}.
A classical example from economics, known as ``Milton Friedman's thermostat'', {describes how a lack of correlation is often confused with a lack of causation in the context of the evaluation of the Federal Reserve} \cite{friedman_feds_2003}. Secondly, often, questions are formulated in terms of (symmetrical) dependence, because it involves established methods and allows a clear interpretation. This could be a case in many risk management applications where the question of what causes losses should be {central}, but is not commonly addressed with causal methods \cite{fenton_risk_2012}. The tools for quantifying causality that are currently being developed can help {to better quantify} causal inference and better understand the results.

In this section, we provide a critique of the methods to help understand their weaknesses and to enable {the reader} to choose the most appropriate method for the intended use. This will also set out the possible directions of future research.
The first part of this section describes the main differences between the methods, followed by a few comments on model selection and problems related to permutation testing. {Suggestions for future research directions conclude the section.

\newpage
\subsection{Theoretical Differences}\index{Geweke's measures}\index{kernelised Geweke's measures}\index{Hilbert--Schmidt Normalised Conditional Independence Criterion (HSNCIC)}\index{transfer entropy}
\label{theo_diff}

Linearity \emph{versus} nonlinearity: The original Granger causality and its Geweke's measure formulation {were} developed to assess linear causality, and they are very robust and efficient in doing so. For data with linear dependence, using linear Granger causality is most likely to be the best choice. The measure can work well also in cases where the dependence is not linear, but has a strong linear component.

As financial data does not normally exhibit stationarity, linearity {or Gaussianity, linear methods should arguably} not be used to analyse them. In practice, requirements on the size of the data {sets} and difficulties with model selection take precedence and mean that linear methods should still be considered.

Direct and indirect causality. 
Granger causality is not transitive, which might be unintuitive. Although transitivity would bring the causality measure closer to the common understanding of the term, it could also make it impossible to distinguish between direct and indirect cause. As a consequence, it could make the measure useless for the purpose of the reduction of dimensionality and repeated information. However, differentiation between direct and indirect causality is not necessarily well defined. This is because adding a conditioning variable can both introduce, as well as remove the dependence between variables \cite{hsiao_autoregressive_1982}. Hence, the notion of direct and indirect causality is relative to the whole information system and can change if we add new variables to the system. Using methods from graphical modelling~\citep{pearl_causality:_2000} could facilitate the defining of the concepts of direct and indirect causality, as these two terms are well defined for causal networks.

Geweke's and kernelised Geweke's measures can distinguish direct and indirect cause in some cases. Following the example of Amblard \citep{amblard_kernelizing_2012}, we suggest comparing the conditional and non-conditional causality measurements as means of distinguishing between direct and indirect cause for both linear and kernel Granger causality.
Measures, like HSNCIC, are explicitly built in such a way that they are conditioned on side information and, therefore, are geared towards picking up only the direct cause; {however, this does not work as intended, as we noticed} that HSNCIC is extremely sensitive to {the} dimensionality of the data. Transfer entropy (in the form we are using) does not consider side information at all. A new measure, called partial transfer entropy \cite{papana_simulation_2013,kugiumtzis_partial_2013}, has been proposed to distinguish between direct and indirect cause.

Spurious causality:\index{spurious causality} Partially covered in the previous point about direct and indirect cause, the problem of spurious causality is a wider one. As already indicated, causality is {inferred in relation} to given data, and introducing more data can both add and remove (spurious) causalities. The additional problem is that data can exhibit many {types of dependency}. None of the methods we {discuss} in this paper is capable of managing several simultaneous {types of dependency}, be it instantaneous coupling, linear or nonlinear causality. We refer the interested reader to the literature on modelling Granger causality and transfer entropy in the frequency domain or using filters \cite{seth_matlab_2010,lungarella_information_2007,dhamala_estimating_2008}.

Numerical estimator: It was already mentioned that Granger causality and kernel Granger causality are robust for small samples and high dimensionality. Both of those measures optimise quadratic cost, which means they can be sensitive to outliers, but the kernelised Geweke's measure can {somewhat mitigate} this with parameter selection. Granger causality for bivariate data has good statistical tests for significance, while the others {do} not and need permutation tests that are computationally expensive. Furthermore, in the case of ridge regression, there is another layer of optimising parameters, which is also computationally expensive. Calculating kernels is also {relatively} computationally expensive (unless the data is high-dimensional), but they are robust for small samples.

{The} HSNCIC is shown to have a good estimator, which, in the limit of infinite data, {does not} depend on the type of kernel. {Transfer entropy, on the other hand, suffers from issues connected to estimating a distribution: problems with a small sample size and high dimensionality. Choosing the right estimator can help reduce the problem. A detailed overview of the possible methods of estimation of entropy can be found in \cite{hlavackova-schindler_causality_2007}. Trentool, one of the more popular open access toolboxes for calculating transfer entropy, uses a nearest neighbour technique to estimate joint and marginal probabilities, that was first proposed by Kraskov \emph{et al}. \cite{kraskov_estimating_2004,lindner_trentool:_2011,vicente_transfer_2011}. The nearest neighbour technique is data efficient, adaptive and has minimal bias \cite{hlavackova-schindler_causality_2007}. The important aspect of this approach is that it depends on a correct choice of embedding parameter and, therefore, does not allow for analysing the information transfer for arbitrary lags. It also involves additional computational cost and might be slower for low dimensional data. We tested Trentool on several data sets and found that the demands on the size of the sample were higher than for the naive histogram and the calculations were slower, with comparable results. The naive histogram, however, does not have good performance for higher dimensions \cite{hlavackova-schindler_causality_2007}, in which case, the nearest neighbour approach would be advised.}

Non-stationarity: This is one of the most important areas for future research. All of the described measures suffer to some degree from an inability to deal with non-stationary data. Granger causality in the original, linear formulation, is the only measure that explicitly assumes stationarity (more precisely, covariance stationarity \cite{granger_investigating_1969,geweke_measures_1984}), and the asymptotic theory is developed for that case. Geweke describes in~\cite{geweke_chapter_1984} special cases of non-stationary processes that can still be analysed within the standard framework and corresponding literature on adapting the linear Granger causality framework to the case of integrated or cointegrated processes \cite{toda_statistical_1995}. In all of those cases, the type of non-stationarity needs to be known, and that is a potential source of new biases \cite{toda_statistical_1995}. The GCCA toolbox
\cite{note8} for calculating Granger causality provides some tools for detecting non-stationarity and, to a limited degree, also for managing it [29]. In the vector autoregressive setting of Granger causality, it is possible to run parametric tests to detect non-stationarity: the ADF test (Augmented Dickey--Fuller) and the KPSS test (Kwiatkowski, Phillips, Schmidt, Shin). For managing non-stationarity, the GCCA toolbox manual \cite{seth_matlab_2010} suggests analysing shorter time series (windowing) and differencing, although both approaches can introduce new problems. It is also advisable to detrend and demean the data, and in the case of economic data, it might also be possible to perform seasonal adjustment.

{The other measures described in this article do not explicitly assume stationarity; however, some assumptions about stationarity are necessary for the methods to work correctly. Schreiber
developed transfer entropy under the assumption that an analysed system can be approximated by stationary Markov processes \citep{schreiber_measuring_2000}. Transfer entropy in practice can be
affected if the time series is highly non-stationary, as the reliability of the estimation of probability densities will be biased \cite{vicente_transfer_2011}, but non-stationarity, due to the
 slow change of the parameters, does not have to be a problem \cite{gomez-herrero_assessing_2010}. For the other two methods, the kernelised Geweke's measure and HSNCIC, the results for estimator
 convergence are available only for stationary data, according to our knowledge. However, the asymptotic results for HSNCIC have been developed for the too restrictive case of i.i.d. data \cite{note9}.
 The results for kernel ridge regression given\linebreak by \cite{hang_fast_????} have been developed for alpha-mixing data.
}

Choice of parameters: Each of the methods requires {parameter selection}; an issue related to model selection described in {Section} \ref{mod:selection}. All of the methods need {a} choice of the number of lags {(lag order)}, while kernel methods additionally require the choice of kernel, kernel parameter (kernel size) and regularisation parameter.

In the case of {the} Gaussian kernel, the effect of the kernel size on the smoothing of the data {can} be understood as follows \cite{kenji_fukumizu_kernel_2007,shawe-taylor_kernel_2004}. The Gaussian kernel $k(x,y) = exp(-{\Vert x - y \Vert ^{2}} / {\sigma ^{2}})$ corresponds to an infinite dimensional feature map consisting of all possible monomials of input features. If we express a kernel as Taylor series expansions, using the basis $1, u, u^{2}, u^{3}, ...$, the random variables, $X$ and $Y$, can be expressed in RKHS by:
\begin{equation}
\begin{split}
\Phi(X) &= k(X,\cdot) \sim (1, c_{1}X, c_{2}X^{2}, c_{3}X^{3},...)^{T}\\
\Phi(Y) &= k(Y,\cdot) \sim (1, c_{1}Y, c_{2}Y^{2}, c_{3}Y^{3},...)^{T}
\end{split}
\end{equation}
therefore, the kernel function can be expressed as follows:
\begin{equation}
k(x,y) = 1+ c_{1}xy + c_{2}x^{2}y^{2} + c_{3}x^{3}y^{3}+...
\end{equation}
and the cross-covariance matrix will contain information on all of the higher-order covariances:
\begin{equation}
\Sigma_{XY} \sim \left( \begin{tabular}{c c c c c }

0 & 0 & 0 & 0 & 0 \\

0 & $c_{1}^{2}Cov[X,Y]$ & $c_{1}c_{2}Cov[X,Y^{2}]$ & $c_{1}c_{3}Cov[X,Y^{3}]$ & ... \\

0 & $c_{2}c_{1}Cov[X^{2},Y]$ & $c_{2}^{2}Cov[X^{2},Y^{2}]$ & $c_{2}c_{3}Cov[X^{2},Y^{3}]$ & ... \\

0 & $c_{3}c_{1}Cov[X^{3},Y]$ & $c_{3}c_{2}Cov[X^{3},Y^{2}]$ &$ c_{3}^{2}Cov[X^{3},Y^{3}]$ & ... \\

0 & ... & ... & ... & ... \\

\end{tabular}
\right)
\end{equation}

According to Fukumizu \textit{et al}. \citep{fukumizu_kernel_2008}, the HSNCIC measure does not depend on the kernel in the limit of infinite data. However, the other parameters still need to be chosen, which is clearly a drawback. The kernelised Geweke's measure optimises parameters explicitly with the cross-validation, while HSNCIC focuses on embedding the distribution in RKHS with any characteristic kernel. Additionally, transfer entropy requires the choice of method for estimating densities and the binning size in the case of the naive histogram approach.

{Another important aspect is the choice of lag order and the number of lags. We observed in Section~\ref{lin:experiment} that the two Geweke's measures were not sensitive to the choice of lags, and we were able to correctly recognise causality both in the case of the smaller and bigger lag {ranges} used. The two other measures, however, behaved differently. HSNCIC is often not able to observe causality in the case of more lags analysed at a time, but performed well for single lags. Transfer entropy flagged spurious causality in one case where the lag was far from the ``true'' one. However, for real data, with a more complex structure, the choice of lag is likely to be important for all measures (see Section \ref{mod:selection})}.

\subsection{Model Selection}
\label{mod:selection}

For the kernel measures, we {observed} that model selection {was} an important issue. In general, the choice of kernel influences the smoothness of the class of functions considered, while the choice of regulariser controls the trade-off between the smoothness of the function and the error of the fit. Underfitting can be a consequence of {a} too large regulariser and {a} too large kernel size (in the case of {a} Gaussian kernel); conversely, overfitting can be a consequence of {a} too small regulariser and {a} too small kernel size. {One of the methods suggested to help with model selection is cross-validation \citep{amblard_kernelizing_2012}. This method is particularly popular and convenient for the selection of kernel size and regulariser in the ridge regression (see Appendix \ref{proced:crossval}). Given non-stationary data, it would seem reasonable to fit the parameters; however, we concluded that cross-validation was too expensive in the computational sense and did not provide the expected benefits.

Another aspect of model selection (and the choice of parameters) is the determination of an appropriate lag order. For kernel methods, increasing the number of lags does not increase the dimensionality of the problem, as could be expected in the case of the methods representing the data explicitly. As described in Section \ref{kernel:Gcausality}\!, in the case of the kernelised Geweke's measure, increasing the number of lags decreases the dimensionality of the problem, due to the fact that the data is represented in terms of $(n-p)\times(n-p)$ pairwise comparisons, where $n$ is the number of observations and $p$ the number of lags. On the other hand, increasing the number of lags will decrease the number of degrees of freedom. This decrease will be less pronounced for kernel methods which allocate smaller weights to higher lags (as is the case in the Gaussian kernel, but not for the linear kernel). Apart from cross-validation, the other approaches to choosing the lag order suggested in the literature are based on the analysis of the autocorrelation function or partial autocorrelation \cite{hamilton_time_1994,lindner_trentool:_2011}.}

We feel that more research is needed on model selection.

\subsection{Testing}

Indications of spurious causality can be generated not only when applying measures of causality, but also when testing those measures. The permutation test that {was} described in Section \ref{permtest} involves {the} destruction of all {types of dependency}, not just causal dependence.
{In practice, this means that, for example, the existence of instantaneous coupling can result in the incorrect deduction of causal inference, if the improvement in prediction due to the existence of causality is confused with improvement, due to instantaneous coupling.} Nevertheless, simplicity is the deciding factor in favour of permutation tests {over} other approaches.

Several authors \cite{seth_assessing_2012,amblard_kernelizing_2012, sun_assessing_2008} propose repeating the permutation test on subsamples to achieve acceptance rates, an approach we do not favour in practical applications. The rationale for using acceptance rates is that the loss {in} significance from decreasing the size of the sample will be more than made up by calculating many permutation tests for many subsamples. We believe this might be reasonable in the case where the initial sample {is} big and the assumption of stationarity {is reasonable}, but that was not the case for our data. We instead decided to report \emph{p}-values for an overlapping running window. This allows us to additionally assess the consistency of the results and does not require {us} to choose the same significance rate for all of the windows.

\subsection{Perspectives}

In the discussion, we highlighted many areas that still require more research. The kernelised Geweke's measure, transfer entropy and HSNCIC detect nonlinear dependence better than the original Granger causality, but {do not} improve on {its other} weakness: non-stationarity. Ridge regression is {a convenient tool in online learning, and it could prove helpful in dealing with non-stationarity \citep{amblard_kernelizing_2012}}. This is clearly an area worth exploring.

\begin{table}[H]
\centering
	\begin{tabular*}{0.965\textwidth}{ll}
\toprule
	\textbf{Measures }			& \textbf{Properties} \\
	\midrule
	    & Linearity \emph{versus} nonlinearity \\
	\cline{2-2}
	Granger causality & assumes linearity; the best method for linear data, the worst for nonlinear \\
	kernelised Geweke's & works for both linear and nonlinear data \\
	transfer entropy & works for both linear and nonlinear data \\
	HSNCIC   & works for both linear and nonlinear data if low dimension \\
	\hline
						& Distinguishing direct from indirect causality \\
	\cline{2-2}
	Granger causality & to some extent by comparing the measure with and without side information \\
	kernelised Geweke's & to some extent by comparing the measure with and without side information\\
	transfer entropy & {not able to (consider partial transfer entropy)}\\
	HSNCIC   & to some extent, as it is designed to condition on side information \\
	\hline
						& Spurious causality \\
	\cline{2-2}
	Granger causality & susceptible \\
	kernelised Geweke's & susceptible \\
	transfer entropy & susceptible \\
	HSNCIC   & susceptible \\
	\hline
						& Good numerical estimator \\
	\cline{2-2}
	Granger causality & yes \\
	kernelised Geweke's & yes \\
	transfer entropy & no \\
	HSNCIC   & yes \\
	\hline
						& Nonstationarity \\
	\cline{2-2}
	Granger causality & v.
	sensitive; test with ADF (Augmented Dickey--Fuller), KPSS (Kwiatkowski, \\& Phillips,  Schmidt, Shin) use windowing, differencing, large lag\\
	kernelised Geweke's & somewhat sensitive; online learning is a {promising} approach \\
	transfer entropy & somewhat sensitive \\
	HSNCIC   & somewhat sensitive \\
	\hline
						& Choice of parameters \\
	\cline{2-2}
	Granger causality & lag\\
	kernelised Geweke's & kernel, kernel size, regularisation parameter, lag; uses cross-validation \\
	transfer entropy & lag, binning size (if histogram approach used) \\
	HSNCIC   & kernel, kernel size, regularisation parameter, lag \\
	\bottomrule
	\end{tabular*}
	\caption{The summary of the main features of the different measures.}
	\label{tab:feat}
\end{table}
Crucially for applications to financial data, more {needs to} be understood about measuring causality in time series with several {types of dependency}. We are not aware of any study that {addresses} this question. We believe this should be approached first by analysing synthetic models. A possible direction of research here is using filtering to prepare data before causal measures are applied. One possibility is frequency-based decomposition. A different type of filtering is decomposition into negative and positive shocks, for example Hatemi-J 
 proposed {an} ``asymmetric causality measure'' based on\linebreak Granger causality~\citep{hatemi-j_asymmetric_2012}.

The third {main direction} of suggested research is building causal networks. There is a substantial body of literature about causal networks for {intervention-based} causality, described in terms of graphical models. Prediction-based causality has been used less often to describe causal networks, but this approach is becoming more popular \citep{green_causality_2003,eichler_granger_2007,amblard_directed_2011,Pozzi13}. Successfully building a complex causal network requires particular attention to side information and the distinction between direct and indirect cause. This is a very interesting area of research with various applications in finance, in particular portfolio diversification, causality arbitrage portfolio, risk management for investments, \emph{etc}.


\section{Conclusions}
\label{conclusions}

We compared causality measures based on methods from the fields of econometrics, machine learning and information theory. After analysing their theoretical properties and the results of the experiments, we conclude that no measure is clearly {superior to} the others. We believe, however, that the kernelised Geweke's measure based on ridge regression is the most practical, performing relatively well for both linear and nonlinear causal structures, as well as for both bivariate and multivariate systems. For the two {real data sets}, we were able to identify causal directions that {demonstrated} some consistency between methods and time windows and that {did not conflict} with the economic rationale. The two experiments {identified} a range of limitations that need to be addressed to allow for a wider application of any the methods to financial data. {Furthermore, neither of the data sets contained higher frequency} data, and working with a high frequency is likely to produce additional complications.

{A separate} question that we only {briefly touched} upon is the relevance and practicality of using any causality measure. This is a question lying largely in the domain of the {philosophy of science}. Ultimately, it is the interpretation of researchers and their confidence in the data that makes it possible to label a relationship as causal rather than only statistically causal. However, while the measures that we analyse cannot discover a true cause or distinguish categorically between true and spurious causality, they can still be very useful in practice.

Granger causality has often been used in economic models and has gained even wider recognition {since} Granger {was awarded the} Nobel prize in 2003. There is little literature on using nonlinear generalisations of the Granger causality in finance or {economics}. We believe that it has great potential, on the one hand, and still many questions to be answered, on the other. While we expect that some of the problems could be addressed with {an} online learning approach and data filtering, more research on dealing with non-stationarity, noisy data and optimal parameter selection is required.


\appendix
\renewcommand\thesubsection{\thesection \Alph{subsection}}
\begin{flushleft}
\textbf{Appendices}
\end{flushleft}
\vspace{-12pt}
\subsection{ Solving Ridge Regression}
\label{solving:rr}

The regularised cost function is Equation~(\ref{eq:reg:costf}):

\begin{equation}
\beta^{\ast} = \operatornamewithlimits{argmin}_{\beta} \dfrac{1}{m} \sum_{i=p+1}^{n}( (w_{i-1}^{i-p})^{T} \beta - x_{i})^2 + \gamma \beta^{T}\beta
\end{equation}

Now solving Equation~(\ref{eq:reg:costf}) gives:
\begin{equation}
\small
\begin{split}
\mathcal{L} &= \dfrac{1}{m} \sum_{i=p+1}^{n} ((w_{i-1}^{i-p})^{T} \beta - x_{i})^2 + \gamma \beta^{T}\beta \\
&= \dfrac{1}{m} (\mathbf{W} \beta - x)^{T}(\mathbf{W} \beta - x) + \gamma \beta^{T}\beta =\\
&= \dfrac{1}{m} (\beta^{T}\mathbf{W}^{T}\mathbf{W} \beta - 2x^{T}\mathbf{W} \beta + x^{T}x) + \gamma \beta^{T}\beta \\
\dfrac{\partial \mathcal{L} }{\partial \beta} &= \dfrac{1}{m} (2\mathbf{W}^{T}\mathbf{W} \beta - \mathbf{W}^{T}x)+ 2\gamma \beta = 0 \Leftrightarrow\\
&\Leftrightarrow \mathbf{W}^{T}\mathbf{W} \beta^{\ast} + \gamma m \beta^{\ast} = \mathbf{W}^{T}x \Leftrightarrow\\
&\Leftrightarrow \beta^{\ast} = (\mathbf{W}^{T}\mathbf{W}+ \gamma m I_{m})^{-1}\mathbf{W}^{T}x
\end{split}
\end{equation}
where $I_{m}$ is an $m \times m$ identity matrix.

The weights, $\beta^{\ast}$, are called the primal solution\index{primal solution}, and the next step is to introduce the dual solution\index{dual solution}~weights.
\begin{equation}
\mathbf{W}^{T}\mathbf{W} \beta^{\ast} + \gamma m \beta^{\ast} = \mathbf{W}^{T}x \Leftrightarrow \beta^{\ast} = \dfrac{1}{\gamma m} \mathbf{W}^{T} (x - \mathbf{W} \beta^{\ast})
\end{equation}
so for some $\alpha^{\ast} \in \mathbb{R}^{n}$, we can write that:
\begin{equation}
\label{eq:primal:dual}
\beta^{\ast} = \mathbf{W}^{T} \alpha^{\ast}
\end{equation}

From the two sets of equation above, we get that:
\begin{equation}
\begin{split}
\alpha^{\ast} &= \dfrac{1}{\gamma m}(x - \mathbf{W} \beta^{\ast}) \\
&\Leftrightarrow \gamma m\alpha^{\ast} = x - \mathbf{W} \beta^{\ast} = x - \mathbf{W}\mathbf{W}^{T}\alpha^{\ast} \Leftrightarrow\\
 &\Leftrightarrow (\mathbf{W}\mathbf{W}^{T} + \gamma m I_{m}) \alpha^{\ast} = x
\end{split}
\end{equation}

This gives the desired form for the dual weights:

\begin{equation}
\label{eq:dualsol}
\alpha^{\ast} = (\mathbf{W}\mathbf{W}^{T} + \gamma m I_{m})^{-1} x
\end{equation}
which depend on the regularisation parameter, $\gamma$.



\subsection{{Background} from Functional Analysis and Hilbert Spaces}
\label{def:fa}

The definitions and theorems below follow \cite{steinwart_support_2008,gretton_measuring_2005,sun_assessing_2008}. All vector spaces will be over $\mathbb{R}$, rather than $\mathbb{C}$; however, they can all be generalised for $\mathbb{C}$ with little modification.

\begin{definition}{(Inner product)} Let $\mathcal{F}$ be a vector space over $\mathbb{R}$. A function $\langle \cdot, \cdot \rangle_{\mathcal{F}} : \mathcal{F} \times \mathcal{F} \rightarrow \mathbb{R}$ is said to be an inner product on $\mathcal{F}$
 if:
\begin{equation}
\begin{split}
\mbox{(i)} \; \; &\langle f_1 + f_2, f \rangle = \langle f_1, f \rangle + \langle f_2, f \rangle, \; \; \; \mbox{for all } f, f_1, f_2 \in \mathcal{F}\\
\mbox{(ii)} \; \; &\langle \alpha f_1,f_2 \rangle = \alpha \langle f_1,f_2 \rangle \; \; \; \mbox{for all } f_1, f_2 \in \mathcal{F}, \alpha \in \mathbb{R}\\
\mbox{(iii)} \; \; &\langle f_1,f_2 \rangle = \langle f_2,f_1 \rangle \; \; \; \mbox{for all } f_1, f_2 \in \mathcal{F}\\
\mbox{{(iv)}} \; \; &\langle f,f \rangle \geq 0 \; \; \; \mbox{and }\langle f,f \rangle = 0 \; \; \; \mbox{if and only if } f = 0
\end{split}
\end{equation}
\end{definition}

\begin{definition}{(Hilbert space)} If $\langle \cdot, \cdot \rangle$ is an inner product on $\mathcal{F}$, the pair $(\mathcal{F}, \langle \cdot, \cdot \rangle )$ is called a Hilbert space if $\mathcal{F}$ with the metric induced by the inner product is complete \cite{note10}.
\end{definition}

One of the fundamental concepts of functional analysis that we will utilise is that of {a} continuous linear operator: for two vector spaces, $\mathcal{F}$ and $\mathcal{G}$, over $\mathbb{R}$, a map $T : \mathcal{F} \rightarrow \mathcal{G}$ is called a (linear) operator if it satisfies $T(\alpha f) = \alpha T(f)$ and $T(f_1 + f_2) = T(f_1) + T(f_2)$ for all $\alpha \in \mathbb{R}, f_1, f_2 \in \mathcal{F}$. Throughout the rest of the paper, we use the standard notational convention $Tf:=T(f)$.

The following three conditions can be proven to be equivalent: (1) linear operator $T$ is continuous; (2) $T$ is continuous at zero; and (3) $T$ is bounded \cite{note11}.
This result, {along} with the Riesz representation theorem given later, is fundamental for the theory of reproducing kernel Hilbert spaces. It should be emphasised that while the operators we use, such as the mean element and cross-covariance operator, are linear, the functions they operate on will not in general be linear. An important special case of linear operators are the linear functionals, which are operators $T : \mathcal{F} \rightarrow \mathbb{R}$.

\begin{theorem}{(Riesz representation theorem)}
In a Hilbert space, $\mathcal{F}$, all continuous linear functionals \cite{note12}
are of the form $\langle \cdot, f \rangle $, for some $f \in \mathcal{F}$.
\end{theorem}

In {Appendix} \ref{solving:rr}, we used the ``kernel trick'' without explaining why it was permissible. The explanation is given below as the {representer theorem}. The theorem {refers} to a loss function, $L(x, y, f(x))$, that describes the cost of the discrepancy between the prediction, $f(x)$, and the observation, $y$, at the point, $x$. {The risk}, $\mathcal{R}_{L,S}$, associated with the loss, $L$, and data sample $S$ is defined {as} the average future loss of the prediction function, $f$.

\begin{theorem}{(Representer theorem)}~\cite{steinwart_support_2008} Let $L : \mathcal{X} \times \mathcal{Y} \times \mathbb{R} \rightarrow [0,\infty )$ be a convex loss, $S := {\{(x_1, y_1), . . . , (x_n, y_n)\} } \in (\mathcal{X} \times \mathcal{Y})^n$ be a set of observations and $\mathcal{R}_{L,S}$ denote associated risk. Furthermore, let $\mathcal{F}$ be an RKHS over $\mathcal{X}$. Then, for all $\lambda > 0$, there exists a unique empirical solution function, which we denote by $f_{S,\lambda} \in \mathcal{F}$, satisfying the equality:
\begin{equation}
\lambda \Vert f_{S,\lambda} \Vert_{\mathcal{F}}^{2} + \mathcal{R}_{L,S} (f_{S,\lambda}) = \inf_{f\in \mathcal{F}} \lambda \Vert f \Vert_{\mathcal{F}}^{2} + \mathcal{R}_{L,S} (f)
\end{equation}
In addition, there exist $\alpha_1, \cdots \alpha_n \in \mathbb{R}$, such that:
\begin{equation}
f_{S,\lambda}(x) = \sum_{i=1}^{n} \alpha_{i} k(x, x_{i}), \; \; \; \mbox{for } x \in \mathcal{X}
\end{equation}
\end{theorem}

Below, we present definitions that are the building blocks of the Hilbert--Schmidt {normalized conditional independence criterion}.

\begin{definition}{(Hilbert--Schmidt norm)}\index{Hilbert--Schmidt norm} Let $\mathcal{F}$ be a reproducing kernel Hilbert space (RKHS) of functions from $\mathcal{X}$ to $\mathbb{R}$, induced by strictly positive kernel $k:\mathcal{X} \times \mathcal{X} \rightarrow \mathbb{R}$. Let $\mathcal{G}$ be an RKHS of functions from $\mathcal{Y}$ to $\mathbb{R}$, induced by strictly positive kernel $l:\mathcal{Y} \times \mathcal{Y} \rightarrow \mathbb{R}$ \cite{note13}.
Denote by $C:\mathcal{G} \rightarrow \mathcal{F}$ a linear operator. The Hilbert--Schmidt norm of the operator, $C$, is defined as
\begin{equation}
\Vert C\Vert_{HS}^{2} := \sum_{i,j} \langle Cv_{i}, u_{j} \rangle_{\mathcal{F}}^{2}
\end{equation}
given that the sum converges, where $u_{i}$ and $ u_{j}$ are orthonormal bases of $\mathcal{F}$ and $\mathcal{G}$, respectively; $\langle v, u \rangle_{\mathcal{F}}, u, v \in \mathcal{F}$ represents an inner product in $ \mathcal{F}$
\end{definition}

Following \cite{gretton_measuring_2005,sun_assessing_2008}, let $\mathcal{H_{W}}$ denote the RKHS induced by a strictly positive kernel $k_{\mathcal{W}} : \mathcal{W} \times \mathcal{W} \rightarrow \mathbb{R}$. Let $X$ be a random variable on $\mathcal{X}$, $Y$ be a random variable on $\mathcal{Y}$ and $(X,Y)$ be a random vector on $\mathcal{X} \times \mathcal{Y}$. We assume $\mathcal{X}$ and $\mathcal{Y}$ are topological spaces, and the measurability is defined with respect to the relevant $\sigma-$fields. The marginal distributions are denoted by $P_{X}, P_{Y}$ and the joint distribution of $(X,Y)$ by $P_{XY}$. The expectations, $\textbf{E}_{X}$, $\textbf{E}_{Y}$ and $\textbf{E}_{XY}$, denote the expectations over $P_{X}$, $P_{Y}$ and $P_{XY}$, respectively. To ensure $\mathcal{H_{X}}, \mathcal{H_{Y}}$ are included in, respectively, $L^{2}(P_{X})$ and $L^{2}(P_{Y})$, we consider only random vectors $(X,Y)$, such that the expectations, $\textbf{E}_{X}[k_{\mathcal{X}}(X,X)]$ and $\textbf{E}_{Y}[k_{\mathcal{Y}}(Y,Y)]$, are finite.


\begin{definition}(Hilbert--Schmidt operator) \index{Hilbert--Schmidt operator}A linear operator $C:\mathcal{G} \rightarrow \mathcal{F}$ is Hilbert--Schmidt if its Hilbert--Schmidt norm exists.
\end{definition}

The set of Hilbert--Schmidt operators $HS(\mathcal{G}, \mathcal{F}) : \mathcal{G} \rightarrow \mathcal{F}$ is a separable Hilbert space with the inner~product:
\begin{equation}
\langle C,D\rangle_{HS} := \sum_{i,j} \langle Cv_{i}, u_{j}\rangle_{\mathcal{F}} \langle Dv_{i}, u_{j}\rangle_{\mathcal{F}}
\end{equation}
where $C,D \in HS(\mathcal{G}, \mathcal{F})$.

\begin{definition}(Tensor product) \index{tensor product} Let $f \in \mathcal{F} $ and $g \in \mathcal{G} $; then, the tensor product operator $f\otimes g : \mathcal{G} \rightarrow \mathcal{F}$ is defined as follows:
\begin{equation}
(f\otimes g)h := f \langle g,h\rangle_{\mathcal{G}},\; \; \; \mbox{for all } h \in \mathcal{G}
\end{equation}
\end{definition}
The definition above makes use of two standard notational abbreviations. The first one concerns omitting brackets when denoting the application of an operator: $(f\otimes g)h$ instead of $(f\otimes g)(h)$. The second one relates to multiplication by {a scalar, and we write} $f \langle g,h\rangle_{\mathcal{G}}$ instead of $f \cdot \langle g,h\rangle_{\mathcal{G}}$.

The Hilbert--Schmidt norm of the tensor product can be calculated as:
\begin{equation}
\begin{split}
\Vert f\otimes g \Vert _{HS}^{2} &= \langle f\otimes g, f\otimes g \rangle_{HS}
= \langle f, (f\otimes g)g \rangle_{\mathcal{F}}\\
&= \langle f, f \rangle_{\mathcal{F}} \langle g,g \rangle_{\mathcal{G}}
= \Vert f \Vert_{\mathcal{F}}^{2} \Vert g \Vert_{\mathcal{G}}^{2}
\end{split}
\end{equation}

When introducing the cross-covariance operator, we will be using the following results for the tensor product. Given a Hilbert--Schmidt operator $L : \mathcal{G} \rightarrow \mathcal{F}$ and $f \in \mathcal{F} $ and $g \in \mathcal{G} $,
\begin{equation}
\label{eq:tens:linop}
\langle L, f\otimes g\rangle_{HS} = \langle f,Lg\rangle_{\mathcal{F}}
\end{equation}

A special case of Equation (\ref{eq:tens:linop}) with the notation as earlier and $u \in \mathcal{F} $ and $v \in \mathcal{G} $,
\begin{equation}
\label{eq:2tens:2linop}
\langle f\otimes g, u\otimes v\rangle_{HS} = \langle f,u\rangle_{\mathcal{F}}\langle g,v\rangle_{\mathcal{G}}
\end{equation}

\begin{definition}{(The mean element)}\index{mean element} Given the notation as above, we define the mean element, $\mu_{X}$, with respect to the probability measure, $P_{X}$, as such an element of the RKHS $\mathcal{H_{X}}$ for which:
\begin{equation}
\langle \mu_{X}, f\rangle_{\mathcal{H_{X}} }:= \textbf{E}_{X}[\langle \phi(X), f \rangle_{\mathcal{H_{X}}}] = \textbf{E}_{X}[f(X)]
\end{equation}
where $\phi:\mathcal{X} \rightarrow \mathcal{H_{X}}$ is a feature map and $f \in \mathcal{H_{X}}$.
\end{definition}

The mean elements exist, as long as their respective norms are bounded, the condition of which will be met if the relevant kernels are bounded.

\subsection{Hilbert--Schmidt Independence Criterion (HSIC)}
\label{HSIC}

As in {Section} \ref{HSNCIC}\!\!\!, following \cite{gretton_measuring_2005,sun_assessing_2008}, let $\mathcal{F_{X}}, \mathcal{F_{Y}}$ denote the RKHS induced by strictly positive kernels $k_{\mathcal{X}} : \mathcal{X} \times \mathcal{X} \rightarrow \mathbb{R}$ and $k_{\mathcal{Y}} : \mathcal{Y} \times \mathcal{Y} \rightarrow \mathbb{R}$. Let $X$ be a random variable on $\mathcal{X}$, $Y$ be a random variable on $\mathcal{Y}$ and $(X,Y)$ be a random vector on $\mathcal{X} \times \mathcal{Y}$. The marginal distributions are denoted by $P_{X}, P_{Y}$ and the joint distribution of $(X,Y)$ by $P_{XY}$.

\begin{definition}{(Hilbert--Schmidt independence criterion (HSIC)} \index{Hilbert--Schmidt Independence Criterion (HSIC)}With the notation for $\mathcal{F_{X}}, \mathcal{F_{Y}}, P_{X}, P_{Y}$ as introduced earlier, we define the Hilbert--Schmidt independence criterion as the squared Hilbert--Schmidt norm of the cross-covariance operator, $\Sigma_{XY}$:
\begin{equation}
\label{eq:HSIC}
HSIC(P_{XY},\mathcal{F_{X}}, \mathcal{F_{Y}}):= \Vert\Sigma_{XY}\Vert_{HS}^{2}
\end{equation}
\end{definition}

We cite without proof the following lemma from \cite{gretton_measuring_2005}:
\begin{lemma}(HSIC in kernel notation)
\begin{equation}
\label{eq:HSICker}
\begin{split}
HSIC(P_{XY},\mathcal{F_{X}}, \mathcal{F_{Y}}):= &\textbf{E}_{X,X',Y,Y'}[k_{\mathcal{X}}(X,X')k_{\mathcal{Y}}(Y,Y')]
 + \textbf{E}_{X,X'}[k_{\mathcal{X}}(X,X')]\textbf{E}_{Y,Y'}[k_{\mathcal{Y}}(Y,Y')]\\
 - &2\textbf{E}_{X,Y}[\textbf{E}_{X'}[k_{\mathcal{X}}(X,X')]\textbf{E}_{Y'}[k_{\mathcal{Y}}(Y,Y')]]
\end{split}
\end{equation}
where $X,X'$ and $Y,Y'$ are independent copies of the same random variable.
\end{lemma}

\subsection{Estimator of HSNCIC}
\label{HSNCIC:est}

Empirical mean elements:
\begin{equation}
\begin{split}
\hat{m}_{X}^{(n)} &= \frac{1}{n} \sum_{i=1}^{n} k_{\mathcal{X}}( \cdot , X_{i})\\ \hat{m}_{Y}^{(n)} &= \frac{1}{n} \sum_{i=1}^{n} k_{\mathcal{Y}}( \cdot , Y_{i})
\end{split}
\end{equation}

Empirical cross-covariance operator:
\begin{equation}
\begin{split}
\hat{\Sigma}_{XY}^{(n)} &= \frac{1}{n} \sum_{i=1}^{n} (k_{\mathcal{Y}}( \cdot , Y_{i}) - \hat{m}_{Y}^{(n)}) \langle k_{\mathcal{X}}( \cdot , X_{i}) - \hat{m}_{X}^{(n)}, \cdot \rangle_{\mathcal{H_{X}}} \\
 &= \frac{1}{n} \sum_{i=1}^{n} \{ k_{\mathcal{Y}}( \cdot , Y_{i}) - \hat{m}_{Y}^{(n)} \} \otimes \{ k_{\mathcal{X}}( \cdot , X_{i}) - \hat{m}_{X}^{(n)} \}
\end{split}
\end{equation}

Empirical normalised cross-covariance operator:
\begin{equation}
\hat{V}_{XY}^{(n)} = (\hat{\Sigma}_{XX}^{(n)} + n \lambda I_{n})^{-1/2} \hat{\Sigma}_{XY}^{(n)} (\hat{\Sigma}_{YY}^{(n)} + n \lambda I_{n})^{-1/2}
\end{equation}
where $n \lambda I_{n}$ is added to ensure {invertibility}.

Empirical normalised conditional cross-covariance operator:
\begin{equation}
\hat{V}_{XY | Z}^{(n)} = \hat{V}_{XY}^{(n)} - \hat{V}_{XZ}^{(n)} \hat{V}_{ZY}^{(n)}
\end{equation}

For $U$ symbolising any of the variables, $(XZ)$, $(YZ)$ or $Z$, we denote by $K_{U}$ a centred Gram matrix, such that each elements equal to: $K_{U, ij} = \langle k_{\mathcal{U}}( \cdot , U_{i}) - \hat{m}_{U}^{(n)}, k_{\mathcal{U}}( \cdot , U_{j}) - \hat{m}_{U}^{(n)} \rangle_{\mathcal{H_{U}}}$; let $R_{U} = K_{U}(K_{U} + n \lambda I)^{-1}$. With this notation, the empirical estimation of HSNCIC can be written as:
\begin{equation}
HSNCIC_{n}:= Tr[R_{(XZ)}R_{(YZ)} - 2R_{(XZ)}R_{(YZ)}R_{Z} + R_{(XZ)}R_{Z}R_{(YZ)}R_{Z}]
\end{equation}

\subsection{Cross-Validation Procedure}
\label{proced:crossval}

Obtaining a kernel (or more precisely, a Gram matrix) is computationally intensive. Performing cross-validation requires the calculation of two kernels (one for testing data, the other for validation data) for each point of the grid. It is most effective to calculate one kernel for testing and validation data at the same time. This is done by ordering the data so that the training data points are subsequent (and the validation points are subsequent), calculating the kernel for the whole (but appropriately ordered) dataset and selecting the appropriate part of the kernel for testing and validation:

\begin{equation}
\bar{K} = \begin{tabular}{|c|c|}
\hline
$K(W_{train},W_{train})$ & $K(W_{train},W_{val})$ \\
\hline
$K(W_{val},W_{train})$ & $K(W_{val}, W_{val})$ \\
\hline
\end{tabular}
\end{equation}

The kernel for the validation point is now a part of the kernel that used both the training and validation~points:
\begin{equation}
\bar{K}_{val} = K(W_{val},W_{train})
\end{equation}

Such an approach is important, because it allows us to use the dual parameters calculated for the testing data without problems with dimensions. Recalling {Equation} (\ref{eq:dualerror}), we can now express the error as:
\begin{equation}
\dfrac{1}{m} (\bar{K}_{val} \alpha^{\ast} - x)^{T}(\bar{K}_{val} \alpha^{\ast} - x)
\end{equation}

Even with an effective way of calculating kernels, cross-validation is still expensive. As described below, to obtain a significance level for a particular measurement of causality, it is necessary to calculate permutation tests and to obtain an acceptance rate or a series of \emph{p}-values for a moving window. In {practice, in order to run several experiments, using a number of measures in a reasonable amount of time, a reasonable compromise is not to perform the cross-validation after every step, but once per experiment and use those parameters in all trials.}

We believe that one of the strengths of the kernelised Geweke's measure, and one of the reasons why kernels are often used for online learning, {lies} in the fact that it is possible to optimise the parameters, but the parameters {do not} have to be optimised each time.

Geweke's measures are based on the optimal linear prediction. While we generalise them to use nonlinear prediction, we can still use the optimal predictor if we employ cross-validation.

In the applications described in the paper, we have used the Gaussian kernel, which is defined as~follows:
\begin{equation}
\label{eq:ker:Gaussian}
k(x,y) = exp(-\dfrac{\Vert x - y \Vert ^{2}}{\sigma ^{2}})
\end{equation}
and the linear kernel {is defined as} $k(x,y) = x^{T}y$.

We use a randomized five-fold cross-validation to choose the optimal parameter, $\gamma$, for the regularisation and the kernel parameters. Let $(x_{t}, y_{t}, z_{t}), t=1, ..., n$ be the time series. We want to calculate $G_{y\rightarrow x\parallel z}$. Based on the given time series, we create a learning set {with} the lag (embedding) equal {to $p$, and }we prepare a learning set following the notation from {Section} \ref{kernel:Gcausality}\!: $(x_{i}, w_{i-1}^{i-p})$, for $i = p+1, ..., n$. The learning set is split randomly into five subsets of equal size. For each $k = 1, ..., 5$, we obtain a  {$k$}
-th validation set and a  {$k$}
-th testing set that contains all data points that {do not} belong to the  {$k$}
-th validation set.

Next, a grid is created, given a range of values for the parameter, $\gamma$, and for the the kernel parameters (the values change in logarithmic scale). For each training set and each point on the grid, we calculate the dual weights, $\alpha \ast$. Those dual weights are used to calculate the validation score---the prediction error for this particular grid point. The five validation scores are averaged to obtain an estimate of the prediction error for each of the points on the grid. We choose those parameters that correspond to the point of the grid with the minimum estimate of the prediction error. Finally, we calculate the prediction error on the whole learning set given the chosen optimal parameters.

As mentioned, the set of parameters from which we choose the optimal one is spread across a logarithmic scale. The whole cross-validation can be relatively expensive computationally, and therefore, an unnecessarily big grid is undesirable.


\acknowledgements{Acknowledgements}

Many thanks to Kacper Chwia\l kowski (University College London, London, UK) for useful discussions and valuable feedback on the manuscript. Special thanks to Maciej Makowski for proofreading and providing general comments to the initial draft. We would also like to express gratitude to Max Lungarella, CTO of Dynamic Devices AG, for making available the code of \citep{lungarella_methods_2007}, which, while not used in any of the experiments described here, {provided} us with important insights into transfer entropy and alternative methods.
Support of the Economic and Social Research Council (ESRC) in funding the Systemic Risk Centre is acknowledged (ES/K002309/1).

\section*{\noindent Author Contributions}

\vspace{12pt}
 {All authors contributed to the conception and design of the study, the collection and analysis of the data and the discussion of the results.}
%


\conflictofinterests{Conflict of Interest}

The authors declare no conflict of interest.

%
%
%
%



\begin{thebibliography}{----}
\bibitem{note5}
\newblock {The use of permutation{, which is more general than rotation,} is helpful when data is short or the analysed time windows are~short.}
\bibitem{note9}
\newblock {We believe that the generalisation from i.i.d. data to alpha-mixing can be done {similarly as} for the HSIC
\cite{chwialkowski_kernel_2014}}
\end{thebibliography}
\bibliographystyle{mdpi}


\end{document}